\documentclass[journal,comsoc]{IEEEtran}
\usepackage[T1]{fontenc}
\usepackage{algorithmic}
\usepackage{algorithm}
\usepackage{subfig}

\ifCLASSINFOpdf
   \usepackage[pdftex]{graphicx}
   \DeclareGraphicsExtensions{.pdf,.jpeg,.png}
\else
\fi
\usepackage{amsmath}
\usepackage[cmintegrals]{newtxmath}
\begin{document}
%
\title{EdgeChain: An Edge-IoT Framework and Prototype Based on Blockchain and Smart Contracts}

%
%
%

\author{
		Jianli~Pan,~\IEEEmembership{Member,~IEEE}
		Jianyu~Wang,~\IEEEmembership{Member,~IEEE}
		Austin~Hester,~\IEEEmembership{Member,~IEEE}
		Ismail~Alqerm,~\IEEEmembership{Member,~IEEE}
		Yuanni~Liu,~\IEEEmembership{Member,~IEEE}
		Ying~Zhao~\IEEEmembership{Member,~IEEE}
\thanks{First manuscript: June 1st, 2018.}
\thanks{J.~Pan, J.~Wang, A. Hester, and I. Alqerm are with the Department of Mathematics and Computer Science in University of Missouri, St. Louis, MO 63121, USA. (Email: {pan, jwgxc, arh5w6, alqermi}@umsl.edu).} 
\thanks{Y. Liu is with the Institute of Future Network Technologies, Chong Qing University of Posts and Telecommunications, China. (Email: liuyn@cqupt.edu.cn).}
}
\maketitle

\begin{abstract}
The emerging Internet of Things (IoT) is facing significant scalability and security challenges. On the one hand, IoT devices are ``weak'' and need external assistance. Edge computing provides a promising direction addressing the deficiency of centralized cloud computing in scaling massive number of devices. On the other hand, IoT devices are also relatively ``vulnerable'' facing malicious hackers due to resource constraints. The emerging blockchain and smart contracts technologies bring a series of new security features for IoT and edge computing. In this paper, to address the challenges, we design and prototype an edge-IoT framework named ``EdgeChain'' based on blockchain and smart contracts. The core idea is to integrate a permissioned blockchain and the internal currency or ``coin'' system to link the edge cloud resource pool with each IoT device' account and resource usage, and hence behavior of the IoT devices. EdgeChain uses a credit-based resource management system to control how much resource IoT devices can obtain from edge servers, based on pre-defined rules on priority, application types and past behaviors. Smart contracts are used to enforce the rules and policies to regulate the IoT device behavior in a non-deniable and automated manner. All the IoT activities and transactions are recorded into blockchain for secure data logging and auditing. We implement an EdgeChain prototype and conduct extensive experiments to evaluate the ideas. The results show that while gaining the security benefits of blockchain and smart contracts, the cost of integrating them into EdgeChain is within a reasonable and acceptable range.

\end{abstract}


%
\IEEEpeerreviewmaketitle

\section{Introduction} \label{sec:intro}

It is predicted that the emerging Internet of Things (IoT) will connect more than 50 billion smart devices by the year 2025~\cite{ERIC20}. It will inevitably change the way we live and work with smart houses, workspaces, transport and even cities on the horizon. However, such trends create significant scalability and security challenges. First, the IoT devices are relatively {\em ``weak''} and most of their data are sent to remote clouds to be processed. Examples include the majority of the smart phones applications and smart home devices such as Google Home and Amazon Echo. But the existing centralized cloud computing model is very difficult to scale with the projected massive number of devices due to the large amount of generated data and the relatively long distance between IoT devices and clouds. Second, the IoT devices are relatively {\em ``vulnerable''} and could be relatively easily controlled by malicious hackers to form ``botnet'' for various attacks~\cite{DDOS162, BOTNET16}. This is aggravated by the fact that most of the cheap IoT devices are with very limited security capabilities, and very poor or even no technical upgrading or maintenance services, though recently Google's Android Things 1.0~\cite{AND18} started pushing this. 

Edge computing~\footnote{Edge computing is also often referred as ``fog computing'', ``Mobile Edge Computing'', or ``Cloudlet'' in different literature, despite slightly different definitions and scopes. We use edge computing or edge cloud in this paper.}~\cite{SHI16,PAN18a,BON12,MEC16,SAT09} is an emerging direction to provide solutions for the IoT scalability issue. It pushes more computing, networking, storage, and intelligence resources closer to the IoT devices, and provide various benefits such as faster response, handling big data, reducing backbone network traffic, and providing edge intelligence. Typical benefited IoT applications include emergency response, augmented reality, video surveillance, speech recognition, computer vision, and self-driving. 

Many works have also been devoted to IoT security. Traditional general-purpose security solutions are not suitable to run on the IoT devices due to the capability constraints~\cite{GAR18}. A typical compromise is to use lightweight IoT security protocols~\cite{LEE14,RFC7252,RAZ12,YAO15,RFC7815,LWIG}. Perimeter based security through firewall~\cite{OPP97,CHE05} does not require running additional software on IoT devices but cannot prevent internal attacks and has been proved ineffective in securing billions of weak devices. Compared with perimeter based trust, zero-trust approaches~\cite{OSB16,WAR14,ZERO16} are proved to be more effective and seem promising. Direct or indirect system-level security approaches, which do not put intensive security-related loads on IoT devices and do not assume the IoT devices being well-maintained, and if enabled with a zero-trust or trustless capabilities, are much needed. Blockchain~\cite{NAK08,NAR16} combined with smart contracts~\cite{SZA96,CHR16} enable a {\em trustless} environment and are recently attracting more attention due to unique features such as data/transactions persistence, tampering resistance, validity, traceability, and distributed fault tolerance. Limited efforts have been made applying them into decentralized IoT and edge computing systems, and two typical work are \emph{Xiong et al.}~\cite{XIO18,XIO17} using game theory and \emph{Chatzopoulos et al.}~\cite{CHA17} focused on computation offloading. In comparison, our research focus is not on consensus mechanism and mining. Instead, we use permissioned blockchain and smart contracts as carrying vehicle, and our major focus is to provision resources for various IoT applications and control and regulate IoT devices' behavior.  

\begin{figure}[!t]
\centering
\includegraphics[width=\linewidth,keepaspectratio=true]{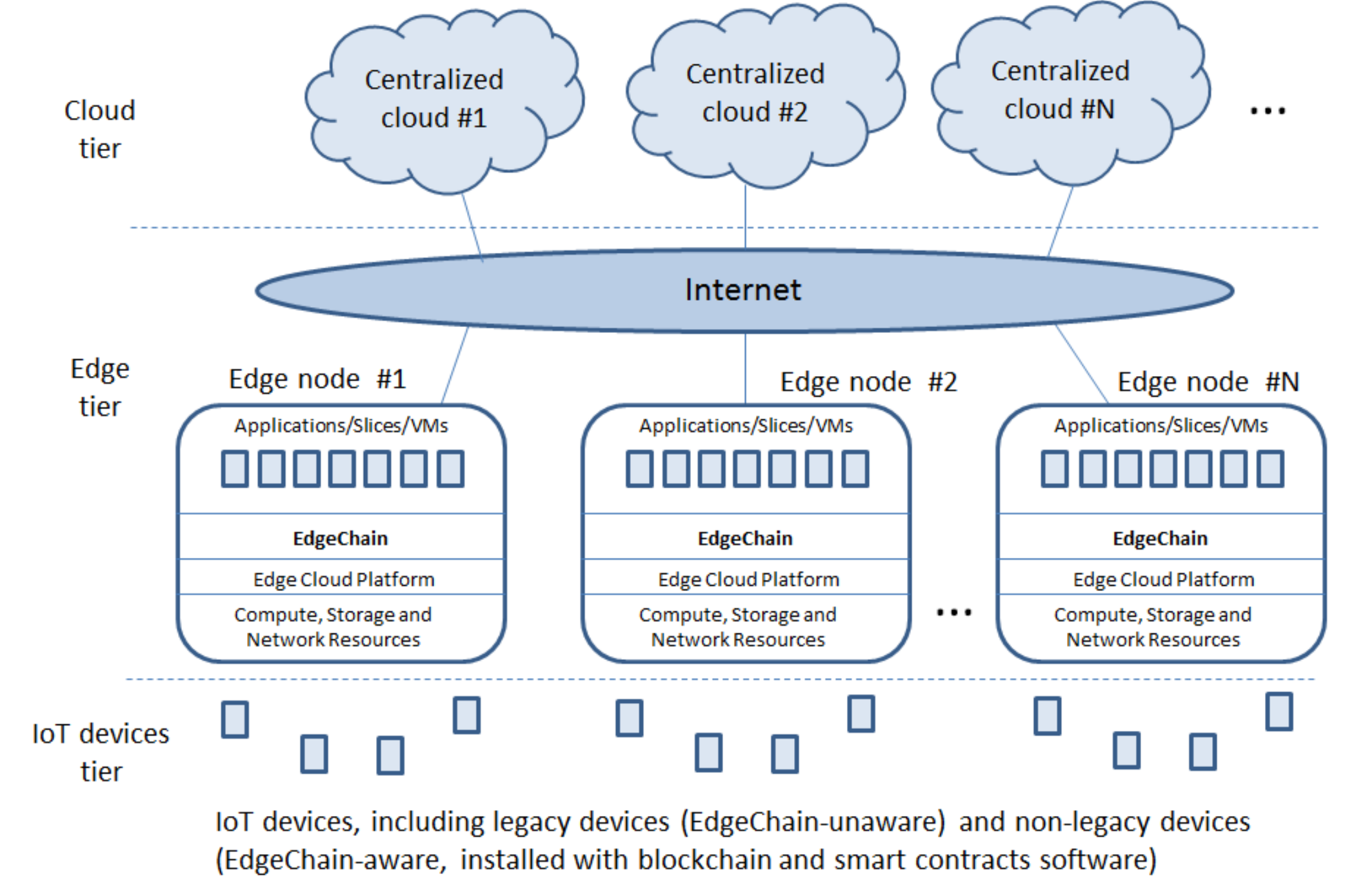}
\caption{EdgeChain position in the multi-tier edge-IoT system network topology.}
\label{fig:position}
\vspace{-10pt}
\end{figure}

In this paper, we seek a fundamentally different approach to tackle these key challenges collectively through a blockchain based and resource oriented edge-IoT framework named \emph{EdgeChain}. The EdgeChain's position in the multi-tier edge-IoT system is illustrated in the Fig.~\ref{fig:position}. As we can see that EdgeChain locates between the edge cloud platforms and the various IoT applications that are launched in the shared infrastructure. It means that EdgeChain can run on different edge cloud platforms such as HomeCloud~\cite{ict16} or Cloudlet~\cite{SAT09}.  

The core EdgeChain idea is to integrate a permissioned blockchain and the internal currency or ``coin'' system to link the edge cloud resource pool with each IoT device' account and resource usage, and hence behavior of the IoT devices. EdgeChain uses a credit-based resource management system to control how much resource IoT devices can obtain from edge servers, based on pre-defined rules on priority, application types and past behavior. Smart contracts are used to enforce the rules and policies to regulate the IoT device behavior in a non-deniable and automated manner. All the IoT activities and transactions are recorded into blockchain for secure data logging and auditing. As a short summary, the major contributions of the EdgeChain framework include:

\begin{enumerate}

\item A new EdgeChain framework integrating permissioned blockchain and smart contracts capabilities.  

\item An internal currency or coin system linking the edge cloud resource pool with IoT device accounts and resource usage behavior.

\item A credit-based resource management system to control how much resources IoT devices can obtain from edge servers.

\item A resource-oritented and smart contracts based policy enforcement method to regulate the IoT device behavior. 

\item A prototype implementation and experimentation to validate and evaluate the EdgeChain ideas. 

\end{enumerate}

Our latest EdgeChain accomplishments have been included in two provisional patents we recently filed~\cite{patent2018a,patent2018b}. Note that EdgeChain is still an ongoing project and some of the work are still in progress. We will discuss the status accordingly in the following sections. The rest of this paper is organized as follows. In Section~\ref{sec:design}, we discuss several key approaches and designs of EdgeChain. We present the EdgeChain framework and functional modules in Section~\ref{sec:framework}. Section~\ref{sec:workflows} is about the key processes and workflows. In Section~\ref{sec:evaluation}, we discuss the prototype and evaluation. Section~\ref{sec:relatedwork} is the related work, while the conclusions and future work follow in Section~\ref{sec:conclusion}.

\section{EdgeChain Key Approaches and Designs} \label{sec:design}
\begin{figure}[!t]
\centering
\includegraphics[width=\linewidth,keepaspectratio=true]{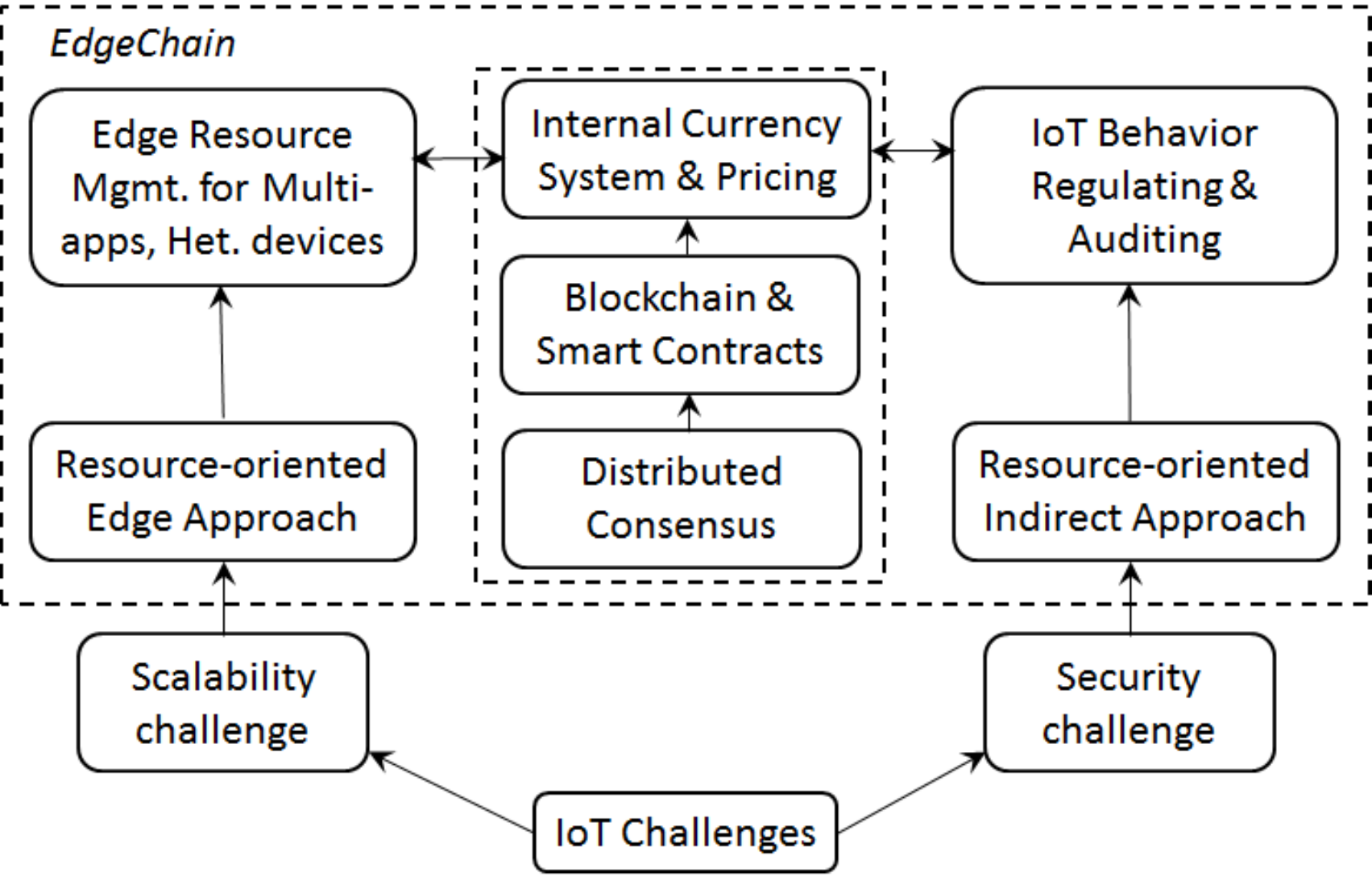}
\caption{EdgeChain vision: the problem space and solution space.}
\label{fig:vision}
\vspace{-10pt}
\end{figure}

In this section, we discuss some key EdgeChain design considerations. Fig.~\ref{fig:vision} shows the overall EdgeChain vision including the problem space and the solution space. 

\subsection{Permissioned Blockchain} \label{subsec:permissioned}
Blockchain networks can be generally categorized into permissionless or public blockchain, and permissioned or private blockchain~\cite{CHR16}. Permissionless blockchain such as Bitcoin network is a peer-to-peer decentralized network. It is usually not controlled by any private organization and the whole network runs on broad consensus of all the members in the network. The trade-off is relatively lower transaction processing throughput and higher latency. Permissioned Blockchain, however, is not a pure peer-to-peer network. The stakeholders such as the application owners of this type of blockchain will have a more controlled and regulated environment, and higer transaction throughput. The consensus mechanisms used for permissionless and permissioned blockchain are also different. 

The EdgeChain system uses a permissoned blockchain since the major goal is to support miscellaneous distributed IoT applications that generally have owners and customers. The system stakeholders need more control and higher throughput and performance. It is not necessary to run very resource-intensive proof-of-work algorithms for consensus and sybil attacks cannot happen. It also removes the necessity of economic incentive for mining, which is usually very resource-consuming in the Bitcoin network. More effective but less resource-intensive consensus protocols are available and a typical example is Practical Byzantine Fault Tolerance (PBFT)~\cite{CAS99} for such an environment.

In EdgeChain, the mining work is only done by the edge servers that have more resources than the IoT devices. It is never done by the resource constrained IoT devices. The mining is much less resource intensive compared with permissionless blockchain network. In other words, the edge servers will be in charge of monitoring the transactions, creating and appending new blocks when new transactions happen. The IoT devices in EdgeChain are only blockchain and smart contracts clients. If they are EdgeChain-aware devices and installed with blockchain and smart contracts software, they are able to interact with the edge servers and get resources and assistance for their tasks through procedures such as cloud offloading~\cite{offloading15}. If they are legacy devices and do not need resources from the edge servers, then they do not even need to install the blockchain and smart contracts software. The EdgeChain is totally transparent to them, but still can create blockchain accounts and manage these IoT devices from the back end. 

\subsection{Credit-based Resource Management}
EdgeChain uses an internal currency or coin system enabled by blockchain to link the edge resource pool with the IoT device accounts and resource usage behavior. EdgeChain consists of a novel credit-based resource management system where each IoT device is created a blockchain account and given an initial amount of credit coins. The credit coin balance determines the device's ability to obtain resources from the edge servers. Generally speaking, the device with a larger balance is afforded quicker and faster access.  The edge server records credits and debits and provides the necessary resources requested by the IoT device based on a set of rules that takes pre-defined priority, application types and past behavior into account. As an ongoing research effort, we are designing detailed intelligent resource provisioning mechanism at the edge clouds for the Quality of Experience (QoE) of multiple applications and heterogeneous devices.  

In fact, we observe that this resource credit management mechanism not necessarily has to be implemented by the internal currency system. The edge server can maintain a traditional credit score system and decide how to grant resources to different devices. However, by utilizing the internal currency system, EdgeChain can gain a series of intrinsic security benefits coming with blockchain. For example, all the coin transactions are automatically logged into the secure and unmodifiable database on blockchain, and it is good for future auditing purposes. Also, it enables smart contracts that could facilitate non-deniable and automated execution of the scheduling rules and policy enforcement in the edge-IoT systems. All these new benefits are not possible without the blockchain and the internal currency system.

\subsection{Resource-oriented, Smart Contracts Based Policy Enforcement and IoT Behavior Regulating}
EdgeChain controls the IoT devices based on their behavior and resource use instead of their locations which results in better security control. This overcomes limitations in existing Edge-IoT solutions which are usually ``perimeter'' based security, i.e., deploying a firewall or a filtering system between the internal and external network and by default trusting the users and nodes ``inside'' the network. If internal IoT devices were hacked and turned to botnet, it is hard to control them.

EdgeChain uses a resource-oriented, smart contracts based, and indirect security scheme for IoT behavior regulating and auditing. EdgeChain adopts an indirect system-level security approach, which means that we do not require the IoT devices to run resource-intensive security software. Instead, EdgeChain monitors, controls and regulates the behavior of IoT devices based on their resource usage and activities. Based on the application types, priority, device's past behavior, the pre-programmed smart contracts enforce the resource policy automatically. It means that if some IoT devices were compromised and controlled by hackers for malicious activities such as behaving erratically, making continuous resource requests that are out-of-line with its profile or application intent, or initiating Denial of Service (DoS) attacks, the smart contracts will execute automatically based on the pre-programmed policies. It will be very soon the device's currency account will run out of balance, through which EdgeChain will be able to quickly identify, control, and contain malicious nodes or devices in the network without requiring them actually to be involved in specific security procedures. EdgeChain can easily take further measures such as putting the devices into the blacklist or blocking the specific devices for further actions. Since smart contracts are based on blockchain, all the activities are recorded into the blockchain. Thus, it is very difficult for any malicious nodes to cause sustained damage or run away with no traces. As an ongoing research effort, we are designing intelligent methods to learn the devices' history behavior pattern based on the data logged in the blockchain to more accurately identify and recognize potential malicious behavior.  

\begin{figure}[!t]
\centering
\includegraphics[width=\linewidth,keepaspectratio=true]{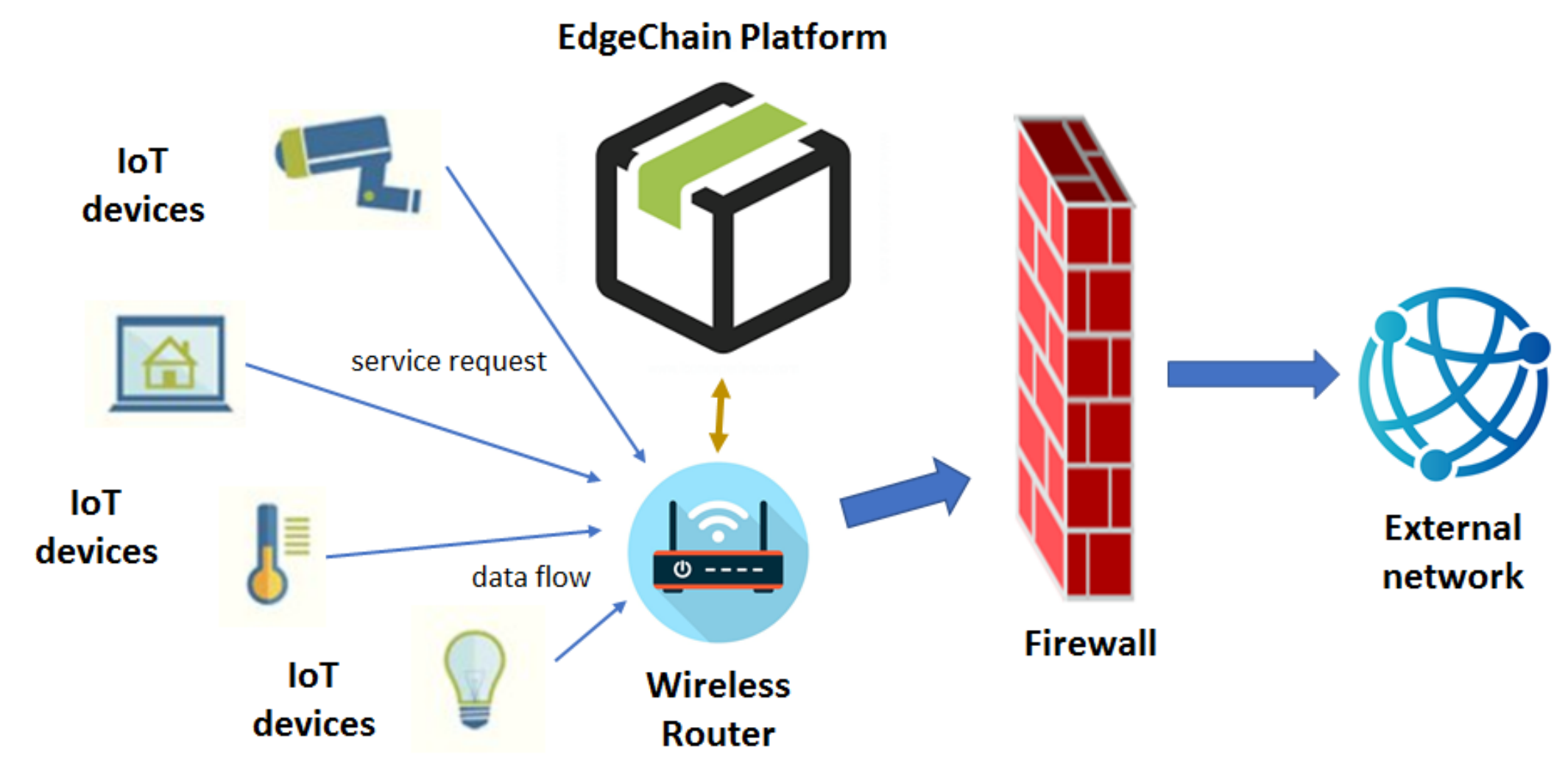}
\caption{A simple example of a standalone EdgeChain box deployment in smart home.}
\label{fig:standalone}
\vspace{-10pt}
\end{figure}

\subsection{Evolutionary and Backward Compatible Approach}  \label{subsec:evolutionary}
We realize the fact that there are a large number of cheap IoT devices that may have very limited security capabilities or are being very poorly maintained and barely upgraded. Though the Google's Android Things 1.0~\cite{AND18} has just been released trying to work on this, it still has a long way to go. There are some extremely incapable IoT devices such as Narrowband IoT (NB-IoT) devices. It may be infeasible to run even the most lightweight blockchain client software. We classify these devices as legacy devices which are EdgeChain-unaware. The other type of devices are relatively capable enough to install with blockchain and smart contracts software and act as a blockchain client. We classify them as non-legacy devices. Non-legacy devices are able to interact with EdgeChain directly and request resources and assistance from the edge servers. Legacy nodes are unaware of the existence of and incapable of working with edge servers. 

The EdgeChain framework adopts an evolutionary and backward compatible approach allowing legacy or extremely incapable IoT devices to work in the new paradigm without assuming them to install new blockchain software or to be updated regularly. The EdgeChain system level capability enables measuring, monitoring, and controlling resource usage of both current and previously installed IoT devices. This goal is achieved through a proxy that works between the legacy IoT devices and the blockchain and smart contract modules, through which the blockchain and smart contracts run transparently to the legacy devices. The proxy sniffs the activities of the legacy nodes and creates blockchain accounts for them just as for non-legacy nodes. In such case, EdgeChain only monitors the behavior and take necessary action if detecting malicious activities. It will not involve allocating edge server resource for the devices. Through the proxy, the legacy IoT devices are not required to know anything about blockchain and smart contracts but they can still be monitored, managed, and controlled by the new Edge-IoT framework. Even if they are compromised by hackers, their malicious behavior can be identified and damages can be contained. 

\subsection{Standalone Deployment vs. Distributed Deployment}
Another important advantage with EdgeChain is the ability to be tailored to the need of the intended application. This allows it to be deployed in both stand-alone modes such as in a smart home as well as distributed modes such as a smart campus or smart city scenario. Fig~\ref{fig:standalone} shows a simple example of a standalone EdgeChain box that is deployed in a smart home. In larger scale use cases and applications such as smart campus and smart cities, multiple such EdgeChain boxes could work in a fully distributed environment, in which cases the distributed boxes work together and share the blockchain and smart contracts data. The edge servers are also able to offload and handover workloads with each other in busy situations. The edge servers can also run appropriate incentive or gaming algorithms associated with their resource pool and blockchain coin accounts to optimize specific goals in revenue, cost, or service latency.

\section{EdgeChain Framework and Functional Modules} \label{sec:framework}

\begin{figure}[!t]
\centering
\includegraphics[width=\linewidth,keepaspectratio=true]{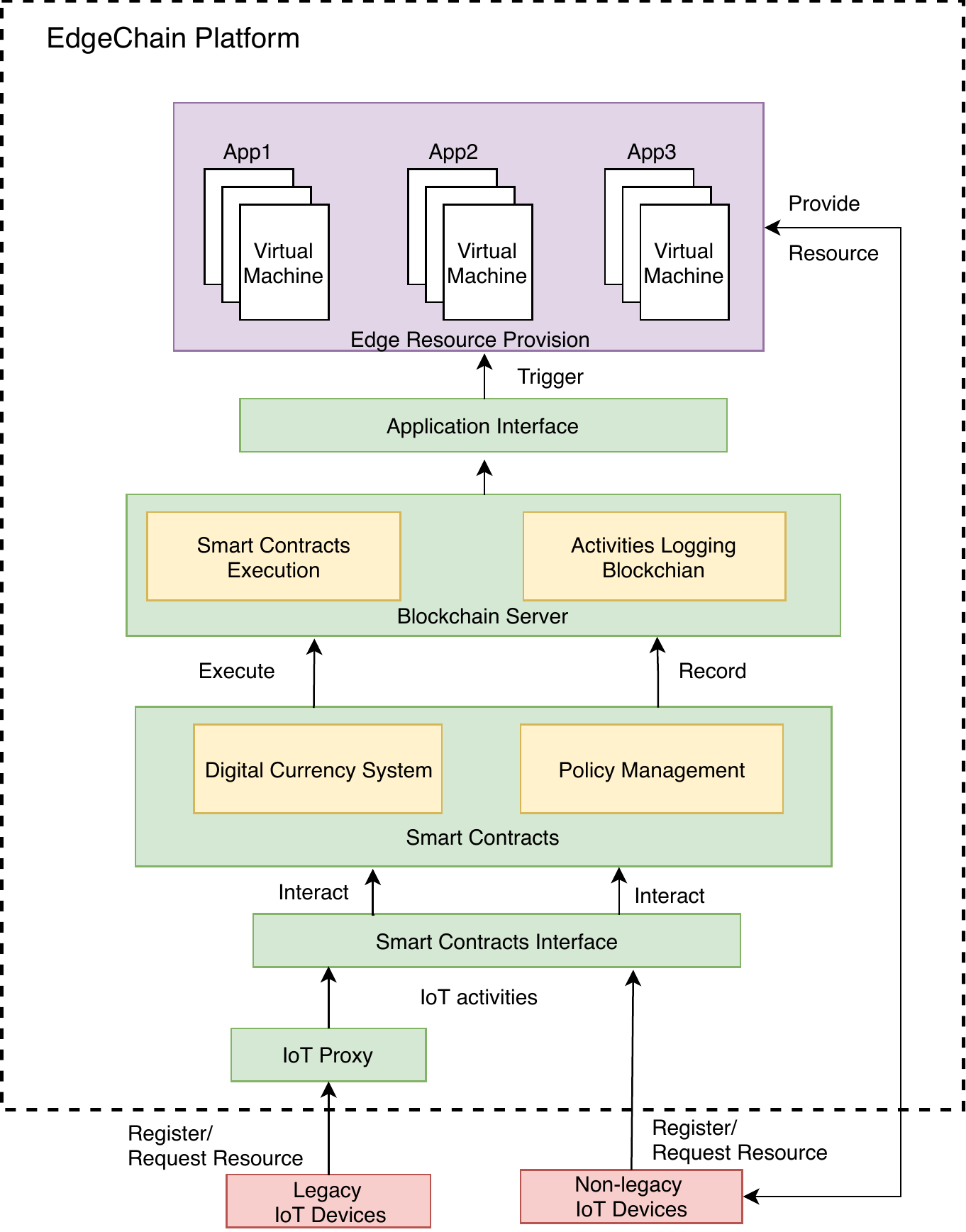}
\caption{EdgeChain framework and functional modules.}
\label{fig:framework}
\vspace{-10pt}
\end{figure}

In this section, we discuss the overall EdgeChain framework and functional modules. The overall system framework is shown in Fig.~\ref{fig:framework}. We can see that the EdgeChain sits between the IoT devices and the edge servers listening to messages and performing corresponding tasks which include device registration and device requests processing. Along the message path, the key modules of EdgeChain include IoT Proxy, Smart Contracts Interface, Smart Contracts, Blockchain Server, and Application Interface. We discuss these modules in a bit more details.  

\subsection{IoT Proxy}

As we discussed in Subsection~\ref{subsec:evolutionary}, the major function of the IoT Proxy module is to accommodate the legacy devices and facilitate their interactions with the blockchain and smart contracts modules. The proxy listens and sniffs the legacy nodes' activities and creates blockchain accounts for them. Registration is done for them in the same way as non-legacy nodes so that the IoT behavior regulating and auditing functions work for them as well. Thus, all their activities are recorded in blockchain as non-legacy nodes. In contrast, the non-legacy devices can interact with smart contract directly and get can get accounts created themselves through the smart contracts interface. Implementing this proxy server function requires appropriate sniffing software and we are currently investigating the most effective open-source tools for the EdgeChain project purposes. 

\subsection{Smart Contracts Interface}
When the IoT activities occur such as registration, communicating between IoT devices, requesting edge server resources, or sending data to outside servers on the Internet, pre-programmed and deployed smart contracts will be triggered to automatically perform corresponding operations and enforce the predefined management rules or policies. Smart Contracts Interface builds a bridge between the IoT applications and the smart contracts. In our implementation, we utilize the Javascript based APIs, named Web3 protocol, to create the smart contract instances for IoT devices. Smart contract instances can call the functions and perform the rules that were encoded in the contracts on behalf of the specific IoT devices.

\subsection{Smart Contracts}
The smart contracts, as the containers of all the rules and policies, consist of two main modules in the EdgeChain system. First, we build a digital currency system whose token are virtual coins to represent the trust levels of IoT devices or their quotas of edge resources they can get. Since every IoT device is bound with a blockchain account, it will be assigned with a certain amount of coins based on its history behavior and resource type. For example, if a device keep behaving well without any malicious actions, it will receive more coins to pay for more service resources. Otherwise, the device may be penalized by being charged more coins to receive the same services or never being rewarded. Second, a module of policy management maintains all the rules that were determined at the time of their creation. The policies can be divided into two types: (1) rules to analyze behavior of IoT devices and handle harmful ones; (2) resource allocation policies to dynamical assign resource to the requests and schedule tasks.

\subsection{Blockchain Server}
In our implementation, the smart contracts are deployed and distributed on the blockchain. The blockchain server provides blockchain service where the IoT devices connect to it as clients. Two functions are performed on the blockchain server. First, the server executes the smart contracts by collecting the transactions among devices and generating the new blocks to run the code embedded in the contracts. Seconds, all the activities in our system are recorded on the blockchain by automatically logging device information, requests and other activities on blocks. This process is also called ``mining'' in the permissonless blockchain. However, as discussed in Subsection~\ref{subsec:permissioned}, the EdgeChain mining process is a lot less resource intensive due to the possible usage of more effective consensus mechanism such as PBFT~\cite{CAS99} and no need for proof-of-work mechanism.

\subsection{Application Interface}
After the interaction with smart contracts and blockchain, there are two possible outcomes: the requests are either rejected due to limited balance in their accounts or malicious behavior identified, or the request are accepted and granted with permission to receive extra edge resource from the edge servers. If the requests are granted, then the IoT devices can interact with the edge server IoT applications directly, e.g., the resource-intensive work such as face recognition from the video stream can be offloaded to the edge servers for faster processing. In this case, Application Interface opens the channels between smart contracts and the edge cloud to trigger resource provision based on the execution results from smart contracts. We achieve this function using Node.js frameworks to listen to the events on the channels and build communications for IoT devices and edge cloud accordingly. 

Note that in terms of delay and time cost, it is true that smart contracts and blockchain operations are not for free and it could take a certain amount of time to finish. The good news is that registration is usually a one-time operation for a specific device. For resource request with the edge servers, after the initial request is granted, the resource provisioning and interactions happen directly between IoT devices and edge servers which will not cause further delay. We conduct very detailed evaluation and experimentation in Section~\ref{sec:evaluation}. 

\subsection{Edge Resource Provisioning}
Once the IoT devices are granted with resources and their accounts are with enough balance for the requested resources, the edge cloud will provision resources in computation, memory, storage, networking, and intelligence accordingly. Since the application may have various requirements for computer capability, bandwidth, latency and privacy, individual virtual machines work as the basic units to meet the specific resource requests. For example, for the video stream based face recognition application example we mentioned, the edge servers could spawn and launch additional virtual machines to process the video stream and get the face recognized. If not sufficient resource available from this edge server, EdgeChain can coordinate with neighbor edge servers to get additional resource. Additional incentive mechanisms and dynamic pricing schemes using game theory or auction can be useful to optimize certain goals in revenue or cost. The IoT devices accounts will be charged accordingly based on the service amount and quality they receive. 

\section{EdgeChain Key Processes and Workflows} \label{sec:workflows}


With all EdgeChain framework and modules, we will discuss the critical processes and workflows in this section. 

\subsection{Blockchain Deployment}

Blockchain implementation can be performed in a distributed way on the edge servers and user devices, and get synchronized across these nodes. We begin by installing blockchain software on the edge server, non-legacy devices, and the IoT Proxy. Our blockchain is built on the Ethereum platform\cite{ethereum14} which is initialized by default to sync with a live public network. However, our EdgeChain system is currently developed for the experimental purpose, so we configure it for use on a private network on campus.

\begin{figure}[!t]
\centering
\includegraphics[scale=0.7]{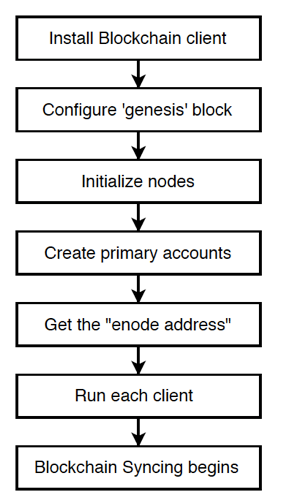}
\caption{Blockchain implementation workflow.}
\label{fig:bcworkflow}
\end{figure}

Fig.~\ref{fig:bcworkflow} shows the workflow of blockchain deployment. The blockchain begins with creating a ``genesis'' block, which holds configuration information such as the hash value of blockhead, timestamp, and difficulty of block mining. It is worth noting that the amount of difficulty makes a significant influence on the mining speed and then on the global system performance since the mining process is realized by solving a Proof-of-Work (PoW) problem with a certain difficulty. Given that only the edge server is permitted to do the mining job, there is no need for a rigorous PoW mechanism to solve the consensus problem. Therefore, our EdgeChain system sets the difficulty to a reasonable low level to balance between over quick mining to avoid storage waste and efficiency of packing transactions. To further reduce the resource consumption of the edge server, we implement an auto-mining function only occurring when there exist unconfirmed transactions.

To sync with one another, all devices must have the same genesis block. The initialization process will provide each node with same genesis configuration. Next, a primary account must be created for each node and public keys are assigned for unique identification. The account gives each node a blockchain address with which it can interact with other nodes and smart contracts. To isolate our system from other public or private blockchains, all nodes are set ``no discovery'' so they cannot connect to other peers without explicit addresses. Such isolation secures the devices from being hooked by external attackers. Thus, each node maintains a specific whitelist called ``enode addresses'' which contains the public keys, IP addresses and network ports of the edge server and some dependent IoT devices. Adding the enode addresses to each node's configuration will allow syncing to occur. Upon completion of the above steps, each node is ready to launch. They will begin seeking friend nodes, syncing and shortly be prepared for use.

\subsection{Development and Deployment of Smart Contracts}

\begin{figure}[!t]
\centering
\includegraphics[scale=0.7]{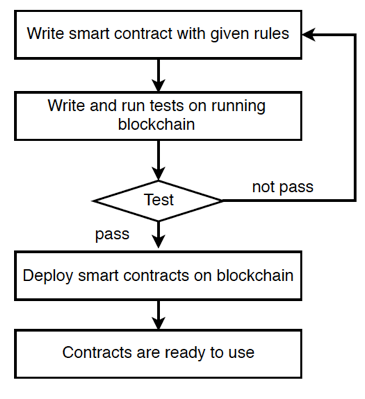}
\caption{Smart contracts implementation workflow.}
\label{fig:scworkflow}
\end{figure}

The proper development of smart contracts guarantees the correct execution of management rules. In our EdgeChain system, the key functional operations including device registration and edge resource allocation are enforced by the corresponding contracts. We deploy smart contracts following the workflow in Fig.~\ref{fig:scworkflow}. When developing a smart contract on the blockchain, it is important to run thorough tests because once deployed, a contract can only be redeployed and lose any data associated with the previous version. Such a redeployment would migrate the contract to the new location and the users may be outdated with an unsupported contract. After deployment, smart contracts are assigned with addresses and treated as normal accounts on blockchain. In order to interact with them, a user must have a copy of the correct address to create an instance as an interface utilizing remote procedure calls (RPC) protocol. The edge server is the performer to execute the functions in the contracts when the IoT devices are the initiators to trigger them.

The smart contracts specify various permissions to different devices where the edge server owns the higher authority to access all the functions but the IoT devices are only limited to some basic functions. Such a setting reduces the impact even if some weak devices are hacked to perform malicious activities. To help engage the legacy nodes into the system, a proxy is deployed in order to fulfill their interaction requests. Other than the direct interaction launched by the nodes, smart contracts are also able to indirectly interface with the outside world by triggering ``events'' which are watched by application interfaces running on the edge server or other nodes on the network. Upon noticing an event of an application, some smart contract can be automatically triggered to execute the predefined tasks. For example, after the edge server finishes serving one user requests, the related service data like service time would be recorded on blockchain by executing a specific contract.

\subsection{Device Registration on Blockchain}

\begin{figure}[!t]
\centering
\includegraphics[scale=0.7]{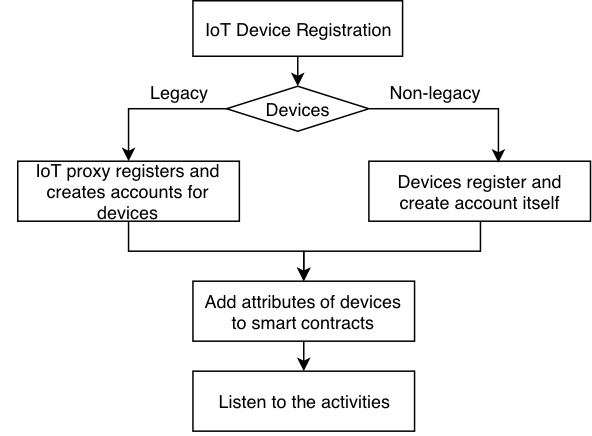}
\caption{Devices registration workflow.}
\label{fig:regworkflow}
\end{figure}

Registration is the first step to engage the IoT devices to be managed and monitored in the EdgeChain system. As illustrated in Fig.~\ref{fig:regworkflow}, the registration starts from determining the type of devices. If there are legacy devices lacking the capability to run blockchain, the proxy can create accounts for each device and register the device specifications stored in the registration smart contracts. If there are non-legacy devices, they can interact with contracts directly to save their attributes by sending transactions.

The registered information makes decisive effect on the request admission introduced in the next section. Specifically, the device specifications partially reference the Manufacturers Usage Description (MUD) \cite{mud18} files which list the activities and communications allowed for IoT devices. Such specifications contain input/output data type, requests of edge resources, MAC address, IP address, network port, communication protocol, and indication flags. Besides, each device registers a unique account address to join blockchain. Upon registration, the edge server will verify the above information and take control of the modification rights of registration data. More parameters will be appended such as priority, coin balance, credit, and requests timestamp to benefit device management. As a summary, Table~\ref{tab: regTable} represents the key device attributes we defined in the registration database which include all the devices key information, value units, and examples. Edge server and IoT devices have different authorities to modify the registry. The attributes marked with ``*'' can only be updated by the edge server. The other basic attributes are filled up during the first registration process initialized by IoT devices.

\begin{table}[!t]
\caption{Registered device attributes.}
\label{tab: regTable}
\centering
\begin{tabular}{|c|c|c|}
\hline
\textbf{Device Specifications} & \textbf{Value Unit} & \textbf{Example}  \\
\hline
account address & string & 0xc968efa8019d (hash value) \\
\hline
network port & int & 42024 \\
\hline
input/output data & string & video,audio,text  \\
\hline
bandwidth request & double & [minValue, maxValue]  \\
\hline
CPU request & double & [minValue, maxValue]  \\
\hline
memory request & double & [minValue, maxValue]  \\
\hline
storage request & double & [minValue, maxValue]  \\
\hline
MAC address & string & 00-14-22-01-23-45  \\
\hline
priority* & int & 1 / 2 / 3 / 4 \\
\hline
coin balance* & double & 200.00 \\
\hline
credit* & int & 100 \\
\hline
isBlocked* & bool & false \\
\hline
isRegistered* & bool & false \\
\hline
last request id* & string & 0xcf30613db6a84 (hash value) \\
\hline
\end{tabular}
\end{table}

\subsection{IoT Behavior Regulation and Activities Management}

The IoT behavior regulation and activities management is the core function of our EdgeChain system for IoT scalability and security. In this subsection, we explain the critical designs in the following order: detailed workflows, edge resource allocation algorithms and behavior management scheme.

\smallskip

\subsubsection{IoT Behavior Regulation Workflow}

EdgeChain not only regulates the activities among IoT devices but also provides the extra edge computing service to boost the resource-intensive applications. When the activities or the requests from IoT devices are recieved, they are treated differently based on the type of devices, either legacy or non-legacy devices. Legacy devices have no request for the support of edge cloud to handle the additional workload. Non-legacy devices could request to obtain edge resource and services under the enforced rules of smart contracts. The detailed workflow is shown in Fig.~\ref{fig:regworkflow} and discussed below.

\begin{figure*}[!t]
\centering
\includegraphics[width=6.5in, height=4.5in]{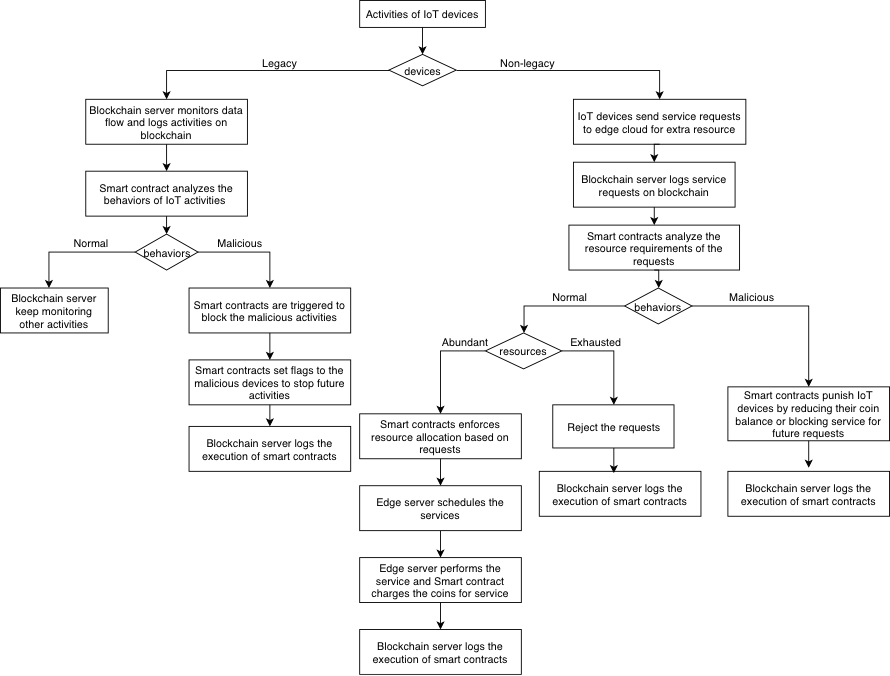}
\caption{IoT activities management workflow.}
\label{fig:actworkflow}
\end{figure*}

For a legacy device, the blockchain server monitors its data flow to other IoT devices or outside network through a sniffer deployed on the IoT gateway such as a WiFi router. During the work process, its activities or behaviors, such as network port and data destination, are logged on blockchain. Then the smart contracts start analyzing the behavior of the device by matching the above observation with the registered attributes. Based on the analyzing results, the blockchain server will choose to keep monitoring the normal behavior. Or it will trigger the smart contract to block any malicious legacy devices and update flags in their registration files. Their future activities will be detected and blocked automatically without performing behavior analysis again. Finally, the execution results of the related smart contracts will be stored on blockchain automatically.

For non-legacy devices, they may send service requests for additional resources for resource-intensive applications such as Virtual Reality (VR) gaming. Once received, the requests are recorded on blockchain in the form of transactions. Next, the resource allocation contracts are executed by the edge server to retrieve the attributes of the devices and analyze the resource requirements in the service requests. If the devices are found to attempt malicious behavior, they will be penalized by reducing their coin balance, lowering credit points and even blocking service for all future requests. If the devices behave normally, the edge cloud will first check the remaining available resource before further process the requests. If the resource pool is exhausted, the requests is rejected and logged. Otherwise, smart contracts perform the resource allocation strategy based on the device types, request details and payable coins. After obtaining the decisions, the edge server starts to schedule the service for the devices immediately. In the meantime, coins will be charged from the devices' account when the edge service begins. Again, the decisions and coin exchanges are all recorded on blockchain. 

\smallskip

\subsubsection{Resource Allocation Based on Pricing Mechanism}

In this specific instance, our optimization goal of resource allocation is to maximize the acceptance rate of user requests. In this case, the currency system plays as the connector among edge server, IoT devices and blockchain by linking edge resource with coins. Our proposed currency system is built on a pricing mechanism to decide: (a) the ordering of the requests may be served; (b) the specific service fee. 

The price of a resource request dynamically changes according to the following environmental parameters: 
\begin{itemize}
  \item Total amount of edge resources
  \item Current available edge resources
  \item Requested edge resource
  \item Application priority
\end{itemize}

Considering the QoE requirements, we categorize the priority of IoT applications into 4 levels, from highest to lowest: (1) Urgent monitoring: patient monitoring, people crowd sensing; (2) Latency sensitive tasks: virtual reality (VR), augmented reality (AR); (3) Reliable data transmission: bank transactions, privacy transferring; (4) Tolerant tasks: light control, sensors based passive monitoring.

\begin{table}[!t]
\caption{Parameters of pricing mechanism.}
\label{tab: priceTable}
\centering
\begin{tabular}{|c|c|c|}
\hline
\textbf{Symbol} & \textbf{Definition} \\
\hline
$N$ & amount of requests at timeslot $t$ \\
\hline
$M$ & number of resource types \\
\hline
$R$ & requested resource $R=\{r_{i\_1},r_{i\_2},...,r_{i\_M} \}$ \\
\hline
$C$ & current availabe resource $C=\{c_{1},c_{2},...,c_{M} \}$ \\
\hline
$W$ & total resources $W=\{w_{1},w_{2},...,w_{M} \}$ \\
\hline
$L$ & priority level \\
\hline
$K$ & amount of accepted requested at timeslot  \\
\hline
$\alpha$ & constant basic price value \\
\hline
$\beta$ & influence factor of priority \\
\hline

\end{tabular}
\end{table}

Table~\ref{tab: priceTable} shows the symbol notations used to calculate the price. We first define the unit price of resource $ j$ for the request $i$:
$$ { P }_{ i\_ j }={ \alpha  }^{ \frac { { r }_{ i\_ j } }{ { c }_{ j } }  }*{ \beta  }^{ { L }_{ i } } $$
Then the total price for request $i$ is defined as, where ${ c }_{ j }\in [0,{ w }_{ j }]$:
$${ P }_{ i }=\sum _{ j }^{ M }{ { { r }_{ i\_ j }*[{ \alpha  }^{ \frac { { r }_{ i\_ j } }{ c\_ j }  }*{ \beta  }^{ { L }_{ i } }] } } =\quad { \beta  }^{ { L }_{ i } }\sum _{ j }^{ M }{ { { r }_{ i\_ j }{ *\alpha  }^{ \frac { { r }_{ i\_ j } }{ c\_ j }  } } }  $$

Using the dynamic pricing, we propose a heuristic request admission algorithm as illustrated in Algorithm~\ref{algPrice}. The proposed algorithm proceeds as follows. At the beginning of timeslot $t$, the number of requests is $N$ and the number of resource types is $M$. For each request ${ r }_{ i\_ j }\in R $, judge if any kind of left edge resource ${ r }_{ i\_ j } $ is less $c_{j}$. If yes, the request is rejected without consideration in this timeslot. If there still have enough resources, calculate the total price of the requests. After all the requests are estimated, the one with the lowest price value is accepted and added to acceptance queue. Then the amounts of available resources $C$ are updated. The rest of requests are reestimated in the next iteration. The algorithm continuous until no request can be admitted. Assume the final acceptance number of request is $K$, we can conclude the time complexity is $O[(N*M+1+M)*K] = O(N*M*K)$, where $K < N$.  Therefore, the algorithm can be solved in polynomial time.

\begin{algorithm}
\small
\caption{Request Admission Algorithm}
\label{algPrice}
\begin{algorithmic}[1]
\REQUIRE $N$ requests $\{req_{1},req_{2},...,req_{N}\}$ at time $t$, request queue $Q(t)$, current available amount of resources $C=\{c_{1},c_{2},...,c_{M} \}$, requested resource $R=\{r_{i\_1},r_{i\_2},...,r_{i\_M} \}$, priority of $req_{i}  L_{i}$ .
\ENSURE accept or deny request $req_{i}$
\item[]
\WHILE {there exists resource for at least one request }
\FOR{each $req_{i}$ in the request queue arrived at timeslot $t$}
\IF{$r_{i\_j}>c_{j}$}
\STATE deny request $req_{i}$;
\STATE continue next iteration;
\ELSIF{$r_{i\_j}<=c_{j}$}
\STATE calculate the total price ${ P }_{ i }$;
\ENDIF
\ENDFOR
\STATE accpet $req_{i}$ with minimal price ${ P }_{ i }$;
\STATE remove the accepted request from $Q(t)$;
\STATE update the avaible edge resources $C$;
\ENDWHILE
\STATE EXIT;
\end{algorithmic}
\end{algorithm}

\smallskip

\subsubsection{Behavior Management Based on Credit System}

Behavior management aims at detecting the potentially malicious activities or requests and taking action to avoid further damage to the system. We propose a credit system to perform the behavior management. Our credit system is distinguished from other similar schemes in the IoT environment because the credit affects resource allocation on the edge server instead of the coorperations between IoT devices. On the other hand, the credit is not directly related to price strategy for edge service but make up the incentive or punishment scheme to restrict the request activities. In this paper, we present the ongoing design and the primary model to show how the credit system works. We consider the following features:

\begin{itemize}
  \item Price threshold: Assume each device only runs one kind of application and sends one kind of resource request, a specific threshold $P_{thres}$  is set for this device i. If $P_{total}$ exceeds $P_{thres}$, the request is regard as potential bad behavior so the deivce credit is reduced. Otherwise, the request is regard as good behavior and credit increases.
  \item Request frequency: If a device continuously send requests in an overhigh frequency, it tends to occupy resource than the common use. So we reduce its credit.
  \item Network port: A device should communicate with the edge server using the predefined network port in the MUD file. Otherwise, some abnormal behavior happens.
  \item Data traffic destination: A device usually has fixed communication targets, so the strange destination indicates the possibility the device is hacked or under control.
\end{itemize}

Each new registered devices owns same initial credits. With the changes of the real-time credit values, we propose two kinds of management actions: (1) If the credit of a device has already been reduced to 0, it is blocked for any future activities; (2) otherwise, the device will get various coin returned based on the credit changes. The equation is defined as follow:
$${ Coin }s_{ return }={ Coins }_{ charged }+\Delta Credit*\eta $$ where $\Delta Credit$ is the change of credit value and $\eta$ is the influence factors of changes. 

We can conclude that the ability to pay for edge service is under the control of the credit system. The better manner receives higher chance to obtain more resources.

\section{Prototype and Evaluation} \label{sec:evaluation}
In this section, we first introduce our experimental testbed built as the EdgeChain prototype. Then, we implement the key functions to verify if it is feasible with acceptable performance overhead. In the third part, two typical IoT applications in different service priorities are deployed on the EdgeChain system to show the compatibility between blockchain and applications. Finally, we test the performance of the pricing-based resource allocation system.

\subsection{EdgeChain Prototype Environment Setup}

\begin{figure}[!t]
\centering
\includegraphics[width=\linewidth,keepaspectratio=true]{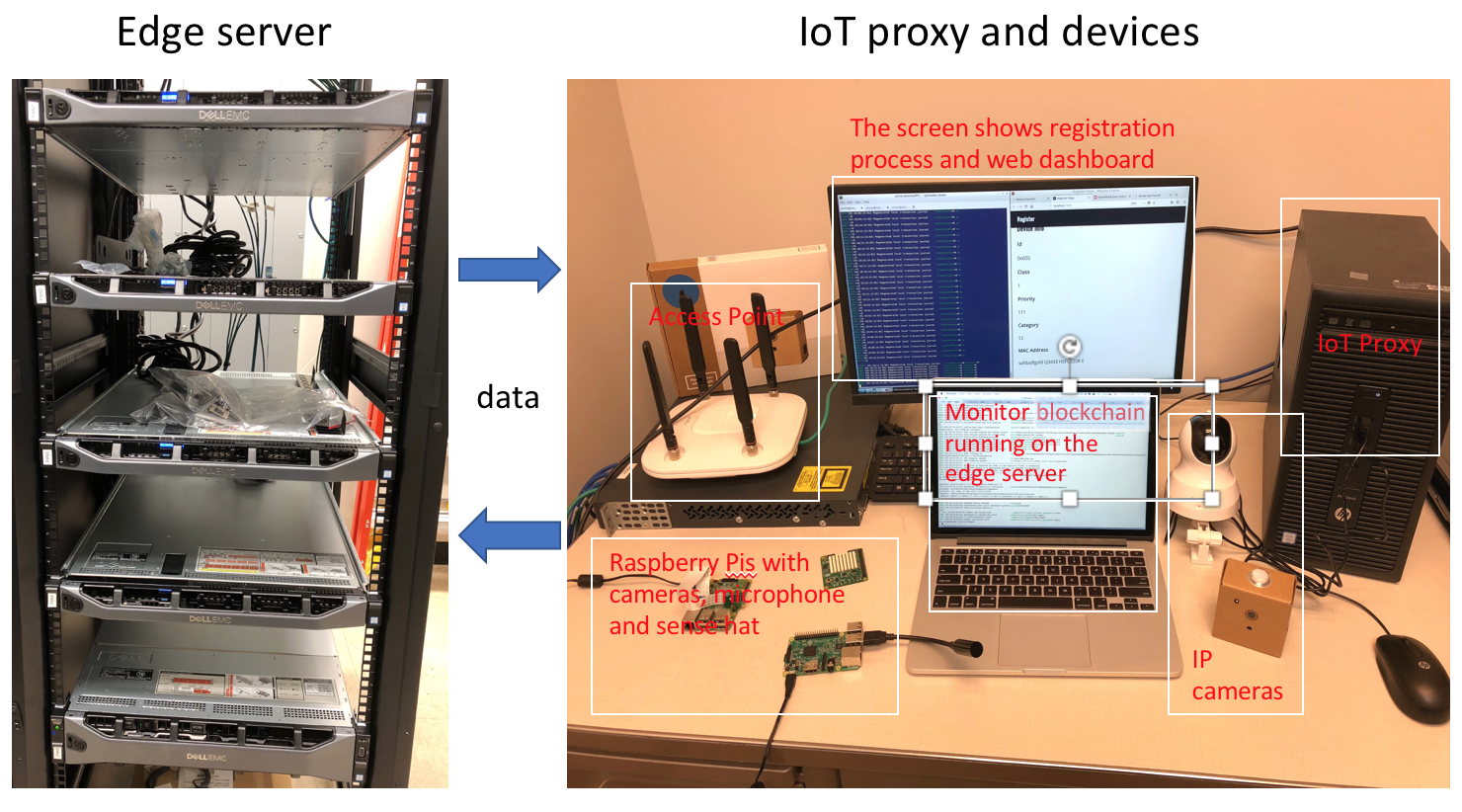}
\caption{EdgeChain testbed.}
\label{fig:testbed}
\end{figure}

The testbed includes the back-end edge cloud cluster and the front-end IoT devices, proxy, and access point. The edge cloud cluster is an OpenStack deployment including 4 high-performance Dell PowerEdge R630 rack servers, 1 high-performance Dell PowerEdge C730x rack server, and 1 high-performance Cisco 3850 switch. The front end consists of several Raspberry Pi 3 Model B single board computers, a Google AIY voice kit, a Google AIY vision kit, and a laptop. One desktop is configured as the proxy for legacy IoT devices, and a high-performance Cisco WiFi Access Point, as illustrated in Fig.~\ref{fig:testbed} 

The detailed hardware and software configurations are as follows. From the aspect of hardware, each OpenStack compute node rack server is equipped with 18 independent CPU cores and 256GB RAM. The mining environment is set up using one core and the rest of the processor cores are reserved for the edge computing service. The miner can boost up to 3.5GHz CPU, 8GB RAM, 1TB storage. As the IoT devices, a Raspberry Pi has 1.2GHz CPU, 1 GB RAM and 32GB storage with several accessory modules including cameras, sense hat, microphone and Google bonnet. The laptop has 2.2GHz CPU, 4GB RAM and 256 GB storage. As for the desktop proxy, 3.2 GHz CPU, 16GB RAM and 1TB storage are installed to manage the multiple blockchain accounts of IoT devices.

Regarding the software, the edge server has installed with \emph{CentOS 7}  as the operating system, \emph{Go-ethereum} as the blockchain running framework, \emph{Solidity} as the smart contract development language, \emph{Truffle} as the contract deployment tool, and \emph{Node.js} as the interface of interactions between IoT applications and blockchain. Except for the blockchain part, the edge computing resources are virtualized using OpenStack cloud platform which helps scale up or down the resource pool flexibly. The edge service is provided in the form of virtual machines to fit the variant specifications of user requests. The Raspberry Pis have been installed with Raspbian operating system and \emph{Go-ethereum} to work in the light mode without block mining function. The laptop is with MacOS and the desktop installs Ubuntu 16.

\begin{figure}[!t]
\centering
\includegraphics[scale = 0.35]{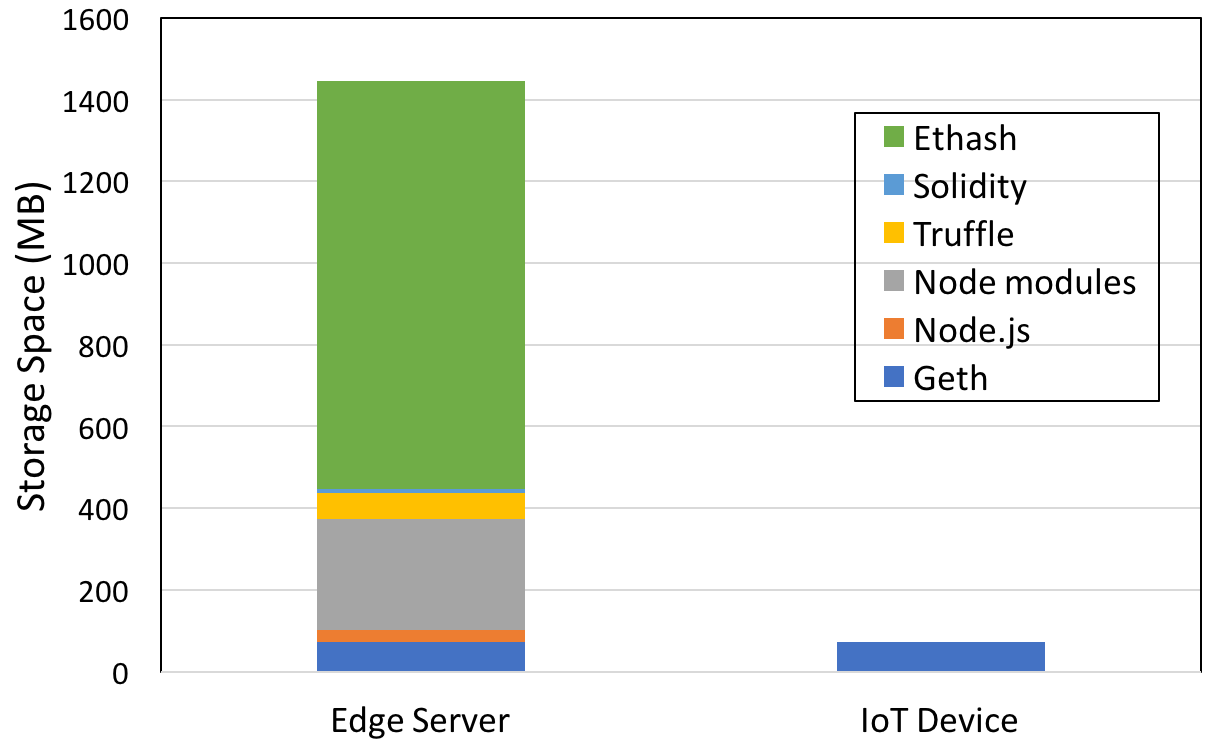}
\caption{Storage of prerequisite software.}
\label{fig:packagesize}
\end{figure}

In the testbed, the rack server works as the edge service provider and the block miner solving Proof-of-Work (PoW) puzzle. The Raspberry Pis and the laptop act as blockchain clients generating and sending transactions of resource requests to the edge server. The desktop interacts with the blockchain on behalf on the legacy devices as a proxy. Given the above installations, the edge server works as a ``full'' blockchain node which stores all the transactions, executes the predefined smart contracts and mines new blocks. The IoT devices work as ``light'' blockchain nodes which only store the transactions data. Fig.~\ref{fig:packagesize} shows the storage requirements for the prerequisites software modules, where \emph{Ethash} is the PoW system used to mine blocks. We put most of the computation work occurring on the blockchain to the full node in order to reduce the overhead on the light nodes.

\subsection{Overhead of Blockchain and Smart Contracts Operation}


We evaluate the blockchain operation based on the two primary functions: IoT devices registration and edge server resource allocation to illustrate the extra overhead caused by the block mining and the interactions of smart contracts. The source of overhead can be divided into three aspects: computation, communication, and storage.

\smallskip

\subsubsection{Computation Cost of Mining Process on Edge Server}

We first evaluate the overhead of device registration in which device specifications are loaded in the transactions signed by their generators. Then the transactions are broadcasted to all the other devices engaged in our system. Finally, these new transactions are packed in the blocks and verified by the miner. We observe the average usage of computation resource on the edge server during mining and no mining, as illustrated in Fig.~\ref{mining}. During the block mining, the edge server consumes much higher CPU and memory resource to commit and packs transactions into new blocks. In contrast, in the idle situation, it only listens to coming transactions such as mining new block caused by new transactions thus consume much less CPU and memory resource.  


\smallskip

\subsubsection{Communication and Storage Cost for Blocks Synchronization}

Given that blockchain is the fully distributed, each device is required to be synchronized with the mainstream chain. The synchronization mechanism relies on the automatic updates and leads to the communication and storage overhead to the system, where the former results from the data transmission and the later from the writing to the local disk. In our system, the edge server maintains the mainstream blockchain and other devices download the chain data from it. In order to evaluate the synchronization delay intuitively, we compare IoT devices to the edge server. Since the edge server as the miner has more computing and bandwidth resource than IoT devices, it completes the validation and transmission of the new blocks faster. As illustrated in Fig.~\ref{miningDelay}, we find the average time to synchronize a new block is 4.09 ms for edge server and 35.9 ms for IoT devices. With higher delay, the IoT devices still meet the latency requirements even for the real-time applications that response time is less than 100ms.

The average size of a block is 128.78 KB and each block can store up to 208 device registrations. Fig.~\ref{blocksize} presents a sample of 50 blocks which have various sizes ranging from 108 KB to 223 KB. Thus, the system will generate around 1.8 MB blockchain data on average for 1,000 devices' registration.


\smallskip

\subsubsection{Computation and Communication Cost of Smart Contract Transactions}

\begin{figure*}[!t]
\centering
\subfloat[Computation resource usage of the edge server for mining.]{\includegraphics[width=2in]{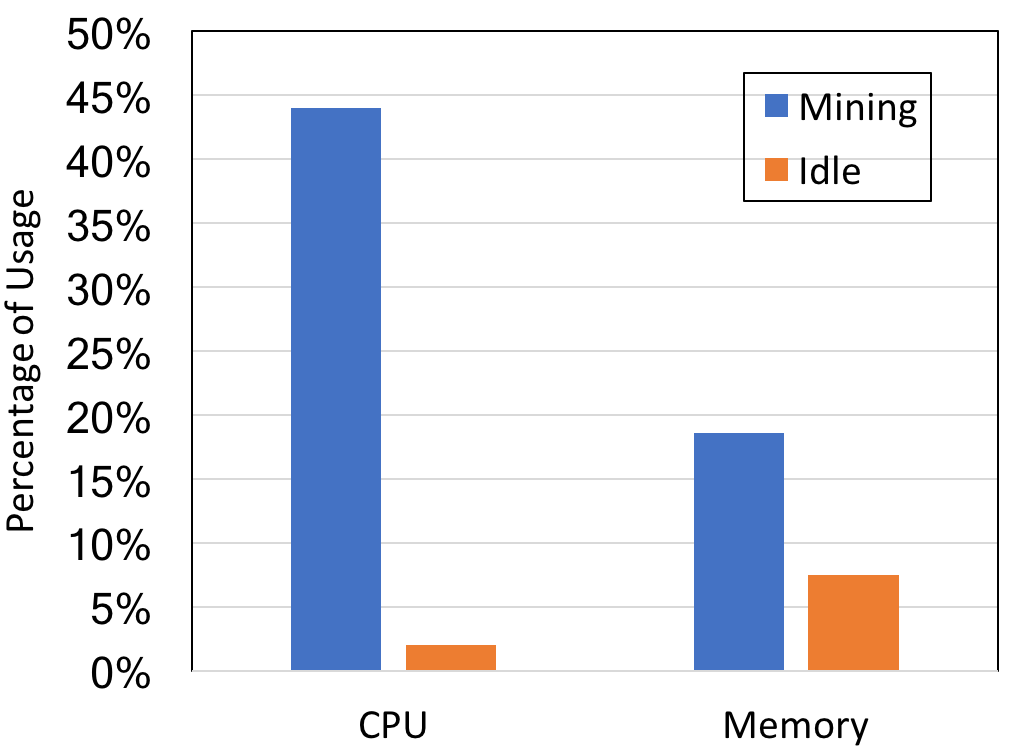}
\label{mining}}
\hfil
\subfloat[Delay to synchronize a block.]{\includegraphics[width=2in]{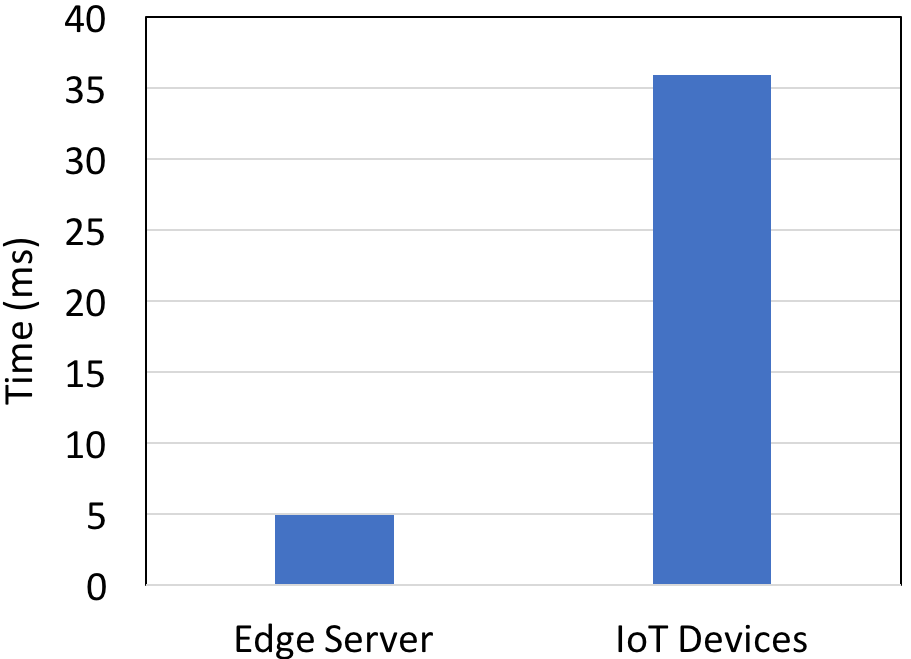}
\label{miningDelay}}
\hfil
\subfloat[Block sizes.]{\includegraphics[width=2in,height=1.5in]{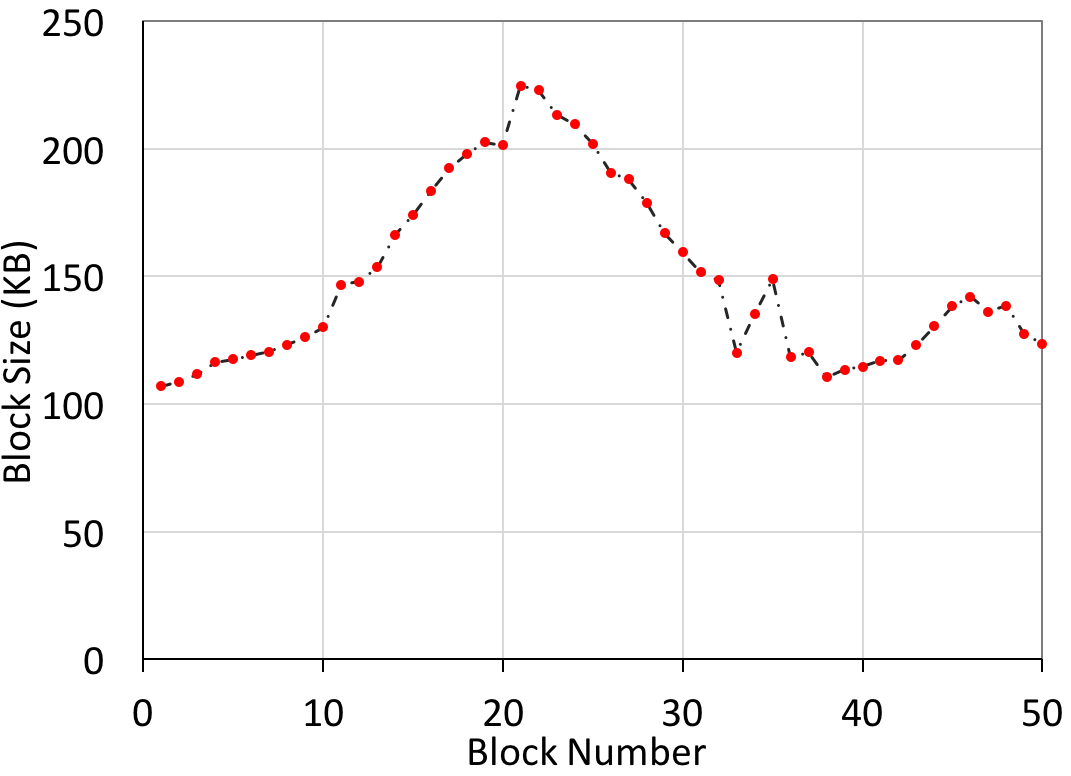}
\label{blocksize}}
\hfil
\subfloat[Computation resource usage for sending transactions.]{\includegraphics[width=2in]{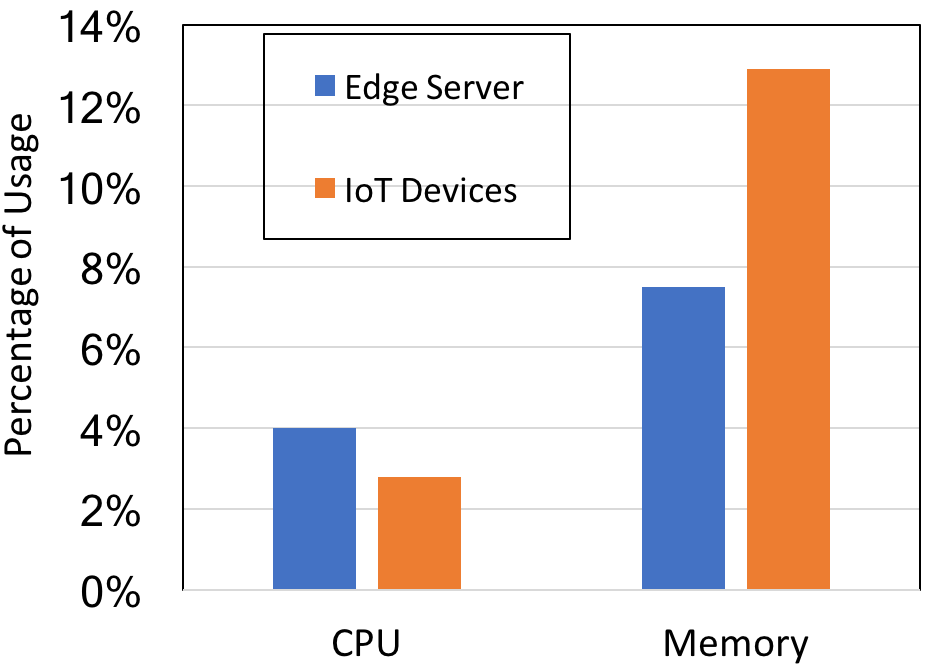}
\label{tranx}}
\hfil
\subfloat[Time to complete one transaction.]{\includegraphics[width=2.2in]{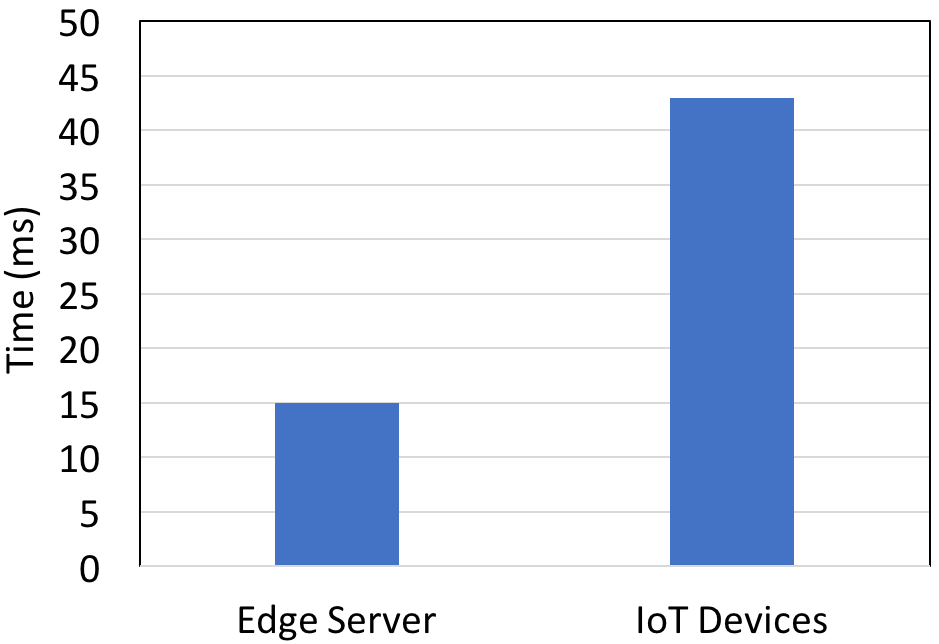}
\label{tranxDelay}}
\caption{Overhead of system operation.}
\label{overhead}
\end{figure*}

In addition to block mining and synchronization, blockchain operation relies on the transactions triggered by the smart contracts. Taking the resource request transaction as an example, we evaluate the computation cost and the interaction delay with smart contracts. The CPU and memory usage are compared between the edge server and IoT devices, as illustrated in Fig.~\ref{tranx}. We observe that the regular transactions take a very low percentage of CPU resource while the memory usage is little higher since the blockchain client occupies 8 \% even in idel time.  We also evaluate the interaction delay of smart contracts which is significant to guarantee system efficiency. Fig.~\ref{tranxDelay} shows the completion of one transaction is less than 50 ms. Such delay should satisfy the latency requirement of the real-time applications.

\subsection{Overhead Comparison of Two Typical IoT Applications}

To evaluate the feasibility and compatibility of the proposed system, we compare blockchain overhead of two typical Edge-IoT applications. We evaluate the face recognition and the natural-language processing applications by testing the computation and communication cost. Face recognition is widely used in the security monitoring applications such as city surveillance, crowd control and door guarding which is latency-sensitive to achieve quick reaction. The typical application of the natural-language processing or voice recognition is the smart home assistant such as Google Home and Amazon echo.

For the face recognition, the Raspberry Pi captures video frames with camera module in 1080p resolution and 60Hz frequency, uploads them to the edge server for image processing and waits for the detection results in the form of location coordinates of detected faces. With regard to the natural-language processing, the Raspberry Pi records the human voice with a USB microphone, transfers it to the edge server, and then the translated text is returned.

\begin{figure*}[!t]
\centering
\subfloat[Comparison of CPU Usage.]{\includegraphics[width=2in]{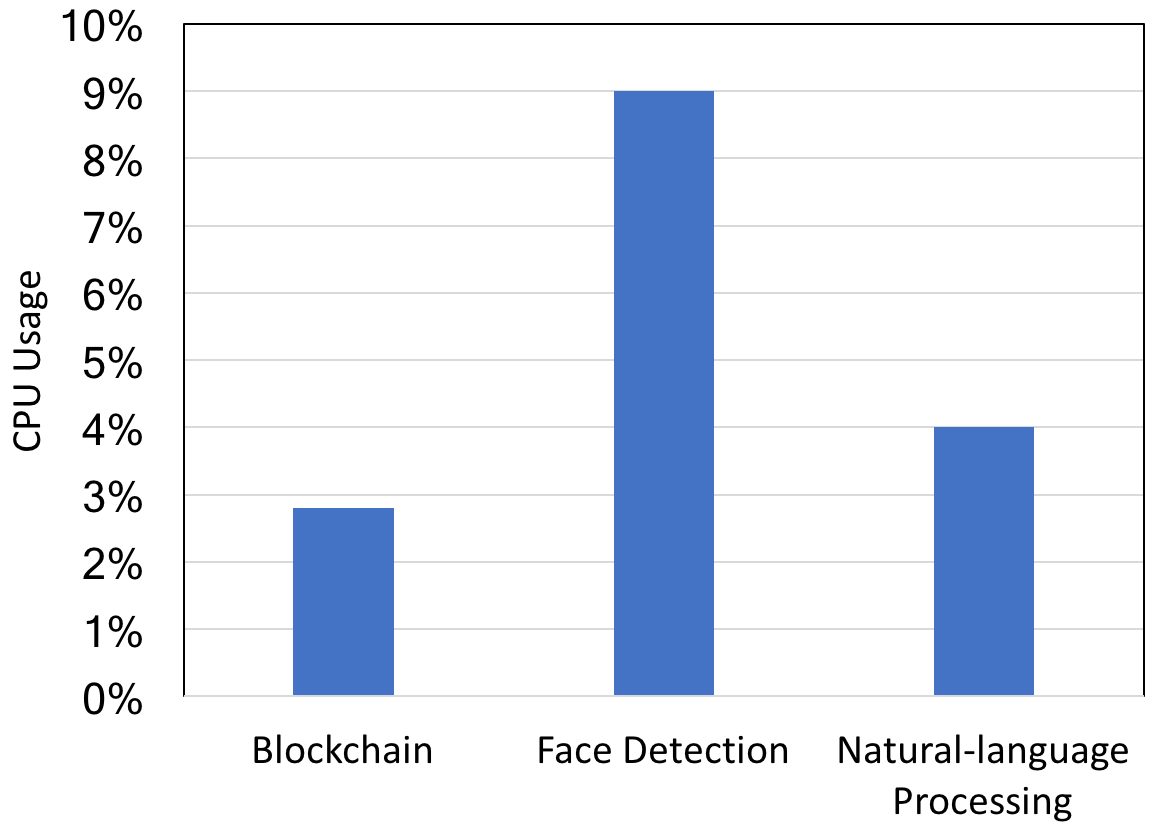}
\label{appCPU}}
\hfil
\subfloat[Comparison of memory Usage.]{\includegraphics[width=2in]{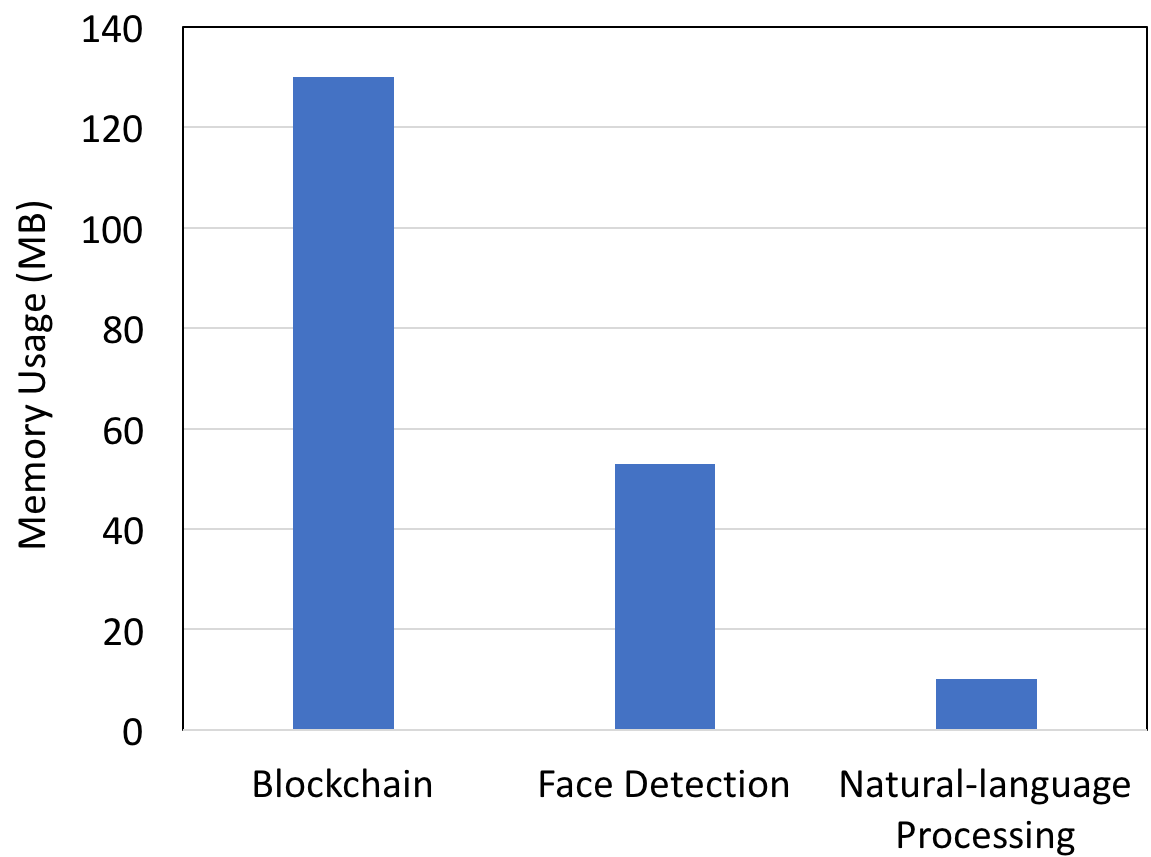}
\label{appMem}}
\caption{Overhead comparison with IoT applications.}
\label{overhead2}
\end{figure*}

We first evaluate the computation cost of blockchain comparing with the two applications. Fig.~\ref{appCPU} shows that the blockchain has the lowest CPU usage compared with the two applications. In addtion, Fig.~\ref{appMem} shows the blockchain has the highest memory usage but still in a low percentage when working with other applications in parallel. Thus, the IoT devices will not suffer from the overload problem. Second, we evaluate the difference of communication data rate among sending blockchain transactions, video and audio data on a Raspberry Pi, as reported in Table~\ref{tab: appComm}. We observe that the regular transactions of resource requests bring very low overhead to the I/O performance and overall network bandwidth.

In summary, we observe that the blockchain can support and collaborate with the IoT application in a distributed and secure way. The overhead is within a reasonable and acceptable range, and the system is feasible to satisfy the requirements to build a multi-application EdgeChain platform for future demands.

\begin{table}[!t]
\caption{Comparison of communication rate}
\label{tab: appComm}
\centering
\begin{tabular}{|c|c|}
\hline
Applications & Data Rate \\
\hline
Blockchain Transactions & 0.54 KB/s\\
\hline
Face Recognition & 1.64 MB/s\\
\hline
Natural-language Processing & 8.12 KB/s\\
\hline
\end{tabular}
\end{table}

\subsection{Resource Allocation Performance of the Pricing Scheme}

\begin{figure*}[!t]
\centering
\subfloat[Influence of $\beta$ value.]{\includegraphics[width=2.3in]{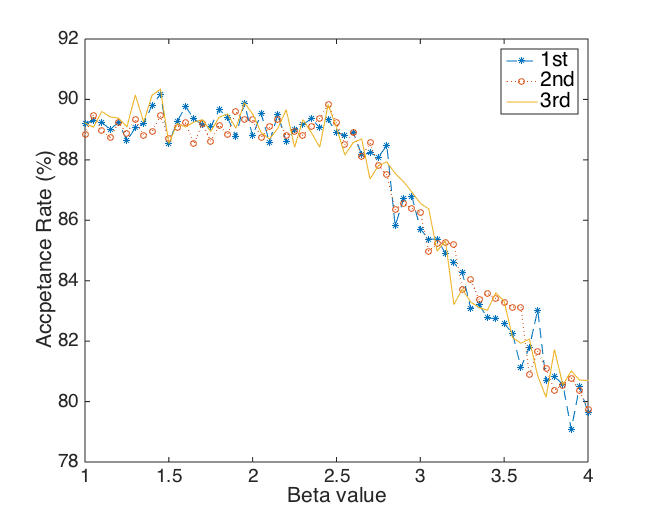}
\label{beta}}
\hfil
\subfloat[Accpentance comparison with constant $\beta$.]{\includegraphics[width=2.2in]{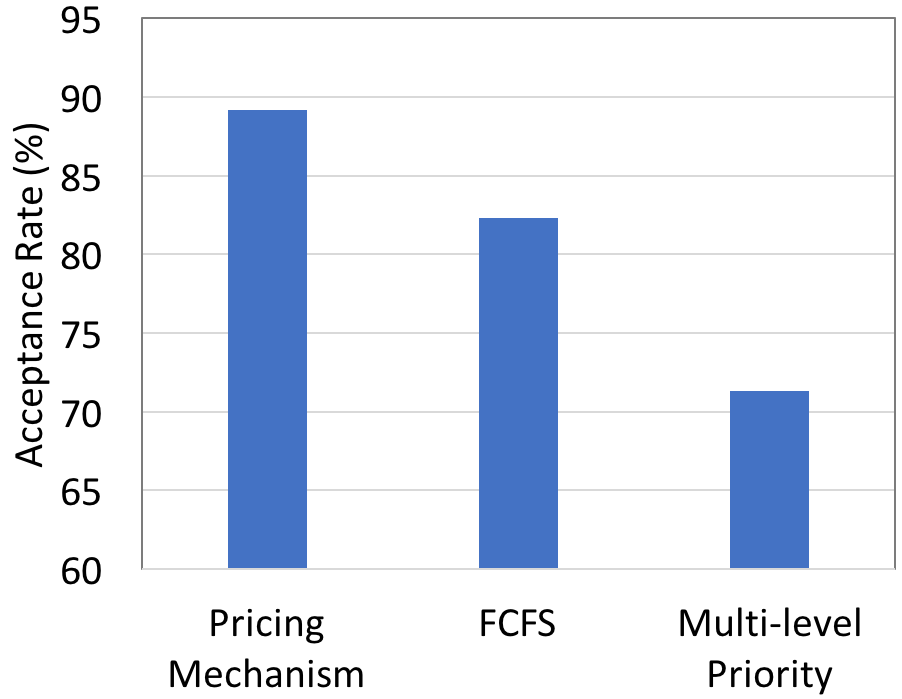}
\label{acRate1}}
\hfil
\subfloat[Accpentance comparison with resource change.]{\includegraphics[width=2.2in]{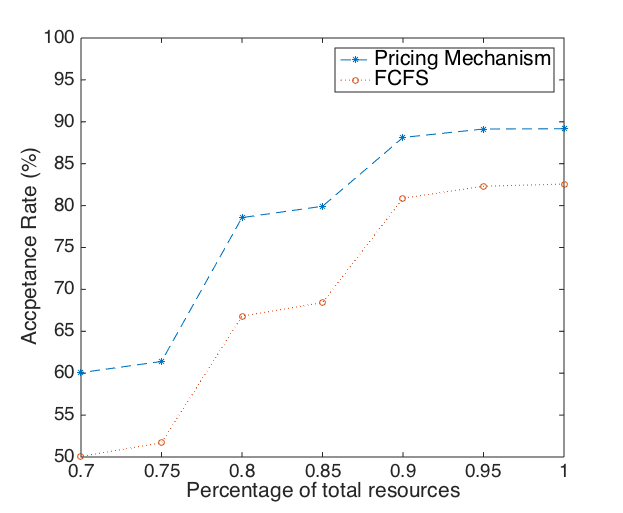}
\label{acRate2}}
\hfil
\caption{Resource allocatioin performance.}
\label{performance}
\end{figure*}

At last, we evaluate the resource allocation performance of the proposed pricing scheme. The goal of resource allocation is to improve the acceptance rate of use requests, which mainly depends on the proposed pricing mechanism. 

We first evaluate the influence of $\alpha$ and $\beta$. $\alpha$ has no effect on the performance since it determines the range of ${ \alpha  }^{ \frac { { r }_{ i\_ j } }{ c\_ j } } $ is located in $[1, \alpha] $. In contrast, $\beta$ adjusts the impact of application priority where the high-priority requests are more likely to be served. We do three random simulations and each one contains 2,000 iteration of random numbers of user requests with different resource requirements. The system parameters are set in Table~\ref{tab: sysConfi} and the request parameters are set in Table ~\ref{tab: reqConfi}. Fig.~\ref{beta} shows the best range of Beta is in [1.3, 1.4] and the too large beta will lead to decrement of acceptance rate since the requests admission simply depends on the priority. 

Second, we compare the acceptance rates among the pricing mechanism, First-Come-First-Serve (FCFS) and multi-level scheduling based on priority, where $\beta=1.35$. Fig.~\ref{acRate1} shows that our proposed algorithm performs best. Then, we evaluate the performance with the change of total edge resources, as illustrated in Fig.~\ref{acRate2}. Starting from the configuration in Table~\ref{tab: sysConfi}, the amount of resources gradually decreases to lower percentages. Our pricing algorithm performs better.

\begin{table}[!t]
\caption{System parameters}
\label{tab: sysConfi}
\centering
\begin{tabular}{|c|c|}
\hline
$\alpha$ & 100 \\
\hline
CPU capacity & 300 \\
\hline
Memory capacity & 250 \\
\hline
Storage capacity & 250 \\
\hline
Bandwidth capacity & 250 \\
\hline
\end{tabular}
\end{table}

\begin{table}[!t]
\caption{Requests parameters}
\label{tab: reqConfi}
\centering
\begin{tabular}{|c|c|c|c|c|c|}
\hline
\textbf{Priority} & \textbf{CPU} & \textbf{Memory} & \textbf{Storage} & \textbf{Bandwidth} & \textbf{Lifetime} \\
\hline
Level 1 & [1,5] & [1,5] & [1,5] & [1,5] & [1,5]  \\
\hline
Level 2 & [10,15] & [5,10] & [5,10] & [1,10] & [1,5]  \\
\hline
Level 3 & [1,5] & [1,5] & [1,5] & [1,5] & [1,5]  \\
\hline
Level 4 & [1,3] & [1,3] & [1,3] & [1,3] & [1,3]  \\
\hline
\end{tabular}
\end{table}

\section{Related Work} \label{sec:relatedwork}
Due to the interdisciplinary essence of EdgeChain, related work comes from different aspects such as IoT, edge computing, blockchain, and smart contracts. A great amount of efforts have been focused on these individual topics, thus, limited by the space, we will not enumerate all the separate efforts. Instead, we will focus on those directly or closely related work. 

The most closely related work are \emph{Xiong et al.}~\cite{XIO18,XIO17} that uses game theory and a pricing mechanism to optimize the profits of the miners at the edge servers. It focuses on the blockchain running costs. \emph{Chatzopoulos et al.}~\cite{CHA17} focuses on computation offloading between devices themselves by using some incentive and reputation schemes. \emph{Sharma et al.}~\cite{SHA17} proposes a conceptual software-defined edge nodes scheme using multi-layer blockchain. Different from these work, our research focus is not on blockchain itself. Instead, we use blockchain as carrying vehicle to provision resources for various IoT applications and control and regulate IoT devices' behavior. More reviewing articles~\cite{CHR16,SUB17,YEO17,DOR16,KSH17} present the overall future prospects in combining blockchain and IoT.

Blockchain and smart contracts are being used to secure many different areas and we will not enumerate them here, but a few example efforts include securing smart home~\cite{DOR17}, securing 5G fog network handover~\cite{SHA18}, securing virtual machine orchestration~\cite{BOZ17}, securing access control in IoT~\cite{ZHA18}, and secure data provenance management~\cite{RAM17}.  

Another thrust of related work is about edge computing research. A large amount of existing work are either on specific applications such as video analytics, vehicular network, cognitive assistance, and emergency response, or very heavily focused on optimizing specific targets such as revenue, cost, delay, or energy consumption associated with operations such as mobile edge offloading, service migration, virtual machines chaining, placement, and orchestration. We will not list all of these works but two good start reading points are~\cite{SHI16,PAN18a}.

%
%
%

\section{Conclusions and Future Work} \label{sec:conclusion}
In this paper, we discussed the design and prototype of the EdgeChain framework which is a novel edge-IoT framework based on blockchain and smart contracts. EdgeChain integrates a permissioned blockchain to link the edge cloud resources with each IoT device's account, resource usage and hence behavior of the IoT device. EdgeChain uses a credit-based resource management system to control the IoT deivces' resource that can be obtained from the edge server. Smart contracts are used to regulate IoT devices' behavior and enforce policies. We implemented an EdgeChain prototype and conducted extensive experiments which showed that the cost for EdgeChain to integrate blockchain and smart contracts are within reasonable range while gaining various intrinsic benefits from blockchain and smart contracts. EdgeChain is still an ongoing project and we are currently working on various issues within the framework such as IoT Proxy, intelligent resource provisioning for multiple heterogeneous applications, and better IoT device behavior regulations.

\section*{Acknowledgment}

The work is supported in part by National Security Agency (NSA) under grants No.: H98230-17-1-0393 and H98230-17-1-0352, and by National Aeronautics and Space Administration (NASA) EPSCoR Missouri RID research grant under No.: NNX15AK38A.

\ifCLASSOPTIONcaptionsoff
  \newpage
\fi




%

%


\begin{IEEEbiography}[{\includegraphics[width=1in,height=1.25in,clip,keepaspectratio]{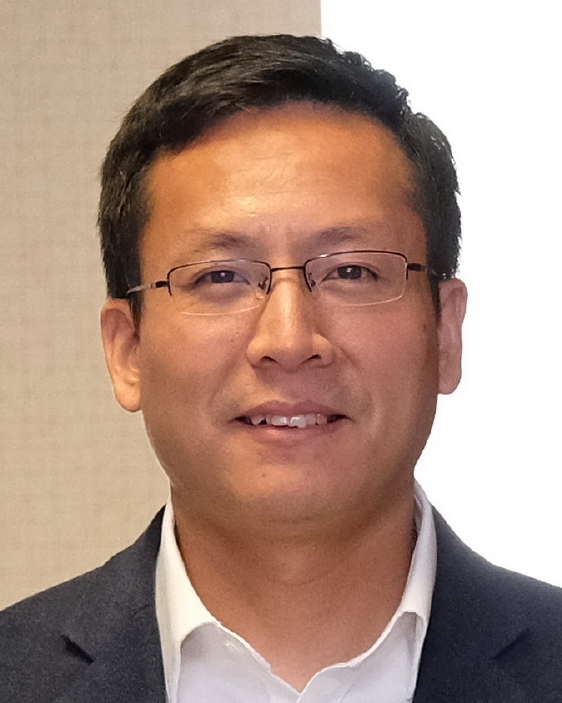}}]{Jianli Pan}

is currently an Assistant Professor in the Department of
Mathematics and Computer Science at the University of Missouri,
St. Louis. He obtained his Ph.D. degree from the Department of
Computer Science and Engineering of Washington University in
St. Louis. He also holds a M.S. degree in Computer Engineering
from Washington University in Saint Louis and a M.S. degree in
Information Engineering from Beijing University of Posts and
Telecommunications (BUPT), China. He is currently an associate
editor for both IEEE Communication Magazine and IEEE Access. His
current research interests include edge clouds, Internet of Things (IoT), Cybersecurity, Network Function Virtualization (NFV), and smart energy. 
\end{IEEEbiography}

\begin{IEEEbiography}[{\includegraphics[width=1in,height=1.25in,clip,keepaspectratio]{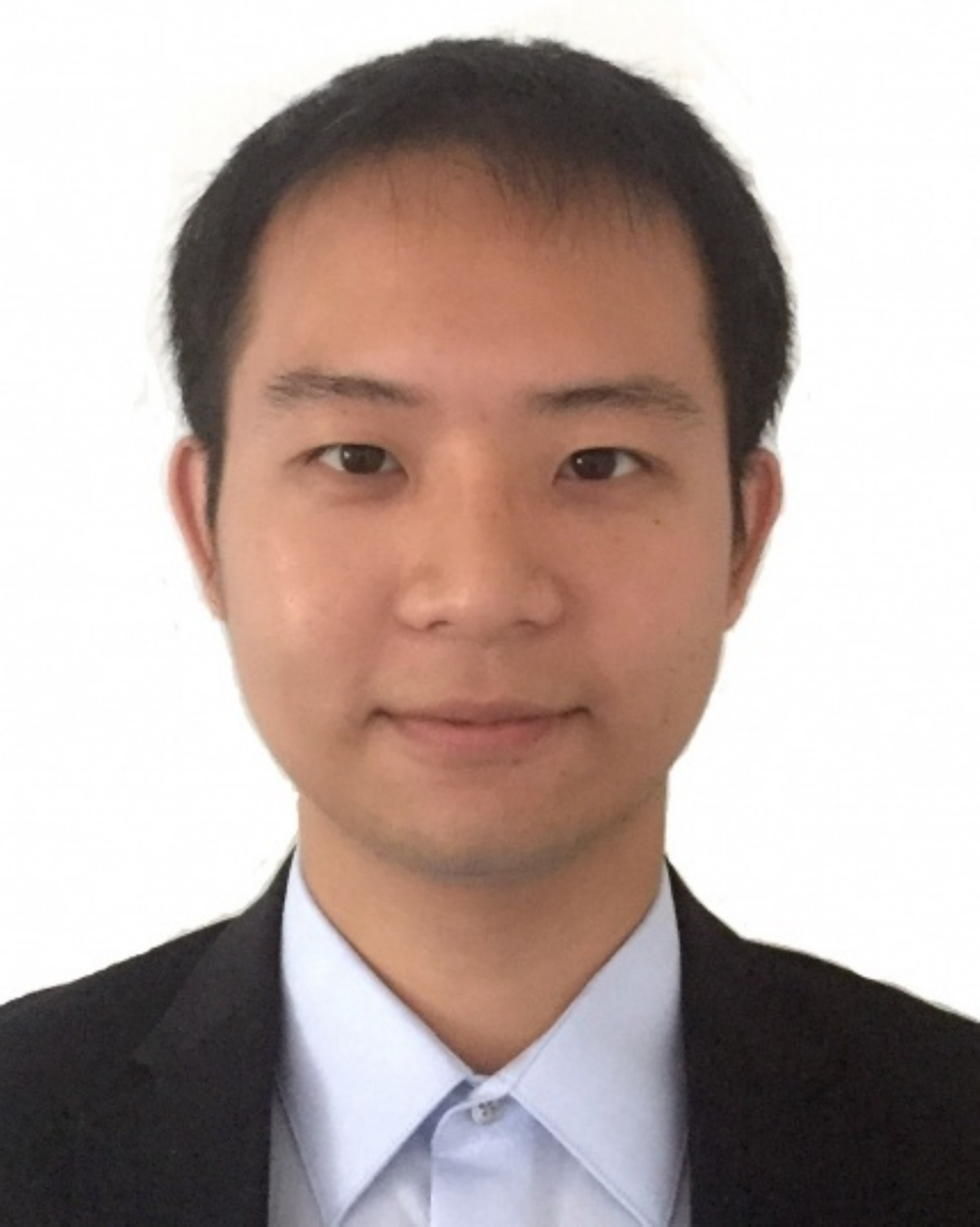}}]{Jianyu Wang}
is currently a Ph.D. student with the Department of Mathematics and Computer Science at the University of Missouri, St. Louis. He received an M.S. in Electrical and Computer Engineering from the Rutgers University, New Brunswick. His current research interests include edge cloud and mobile cloud computing. 
\end{IEEEbiography}

\begin{IEEEbiography}[{\includegraphics[width=1in,height=1.25in,clip,keepaspectratio]{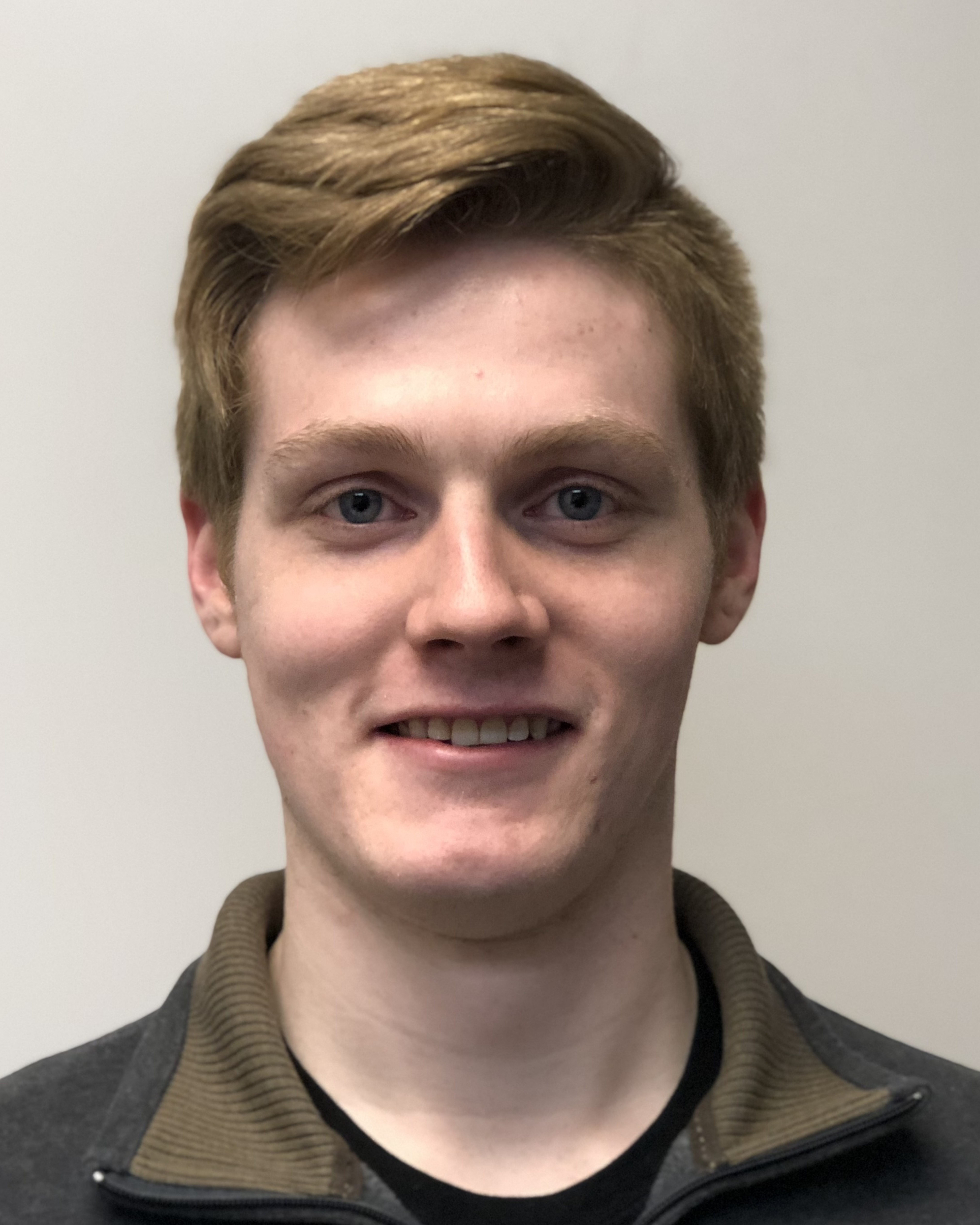}}]{Austin Hester}
is currently an undergraduate student with the Department of Mathematics and Computer Science at the University of Missouri, St. Louis. His current research interests include Internet of Things and Blockchain.
\end{IEEEbiography}

\begin{IEEEbiography}[{\includegraphics[width=1in,height=1.25in,clip,keepaspectratio]{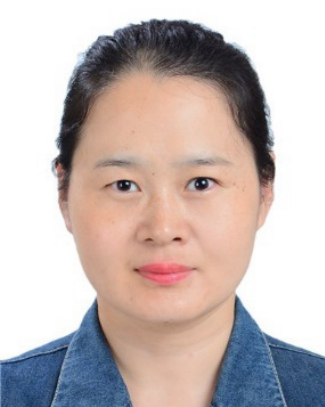}}]{Yuanni Liu}
is an associate professor at the Institute of Future Network Technologies, Chong Qing University of Posts and Telecommunications. She received her Ph.D. from the Department of network technology Institute, Beijing University of Posts and Telecommunications, China, in 2011. Her research interests include mobile crowd sensing, IoT security, and data virtualization.
\end{IEEEbiography}




\vfill


\end{document}